# Superconducting transition temperatures in electronic and magnetic phase diagram of a superconductor, $Sr_2VFeAsO_{3-\delta}$

Yujiro Tojo[1], Taizo Shibuya[2*], Tetsuro Nakamura[1], Koichiro Shoji[1], Hirotaka Fujioka[1], Masanori Matoba[1], Shintaro Yasui[3], Mitsuru Itoh[3], Soshi Iimura[3,4], Hidenori Hiramatsu[3,4], Hideo Hosono[3,4], Shigeto Hirai[5,6**], Wendy Mao[5,6], Shinji Kitao[7], Makoto Seto[7], and Yoichi Kamihara[1]

1)*Department of Applied Physics and Physico-Informatics, Keio University, Yokohama 223-8522,Japan*
2) *Department of Mechanical Engineering, Keio University, Yokohama 223-8522, Japan*
3) *Laboratory for Materials and Structures, Tokyo Institute of Technology,4259 Nagatsuta, Yokohama 226-8503, Japan*
4) *Materials Research Center for Element Strategy, Tokyo Institute of Technology,4259 Nagatsuta, Midori-ku, Yokohama 226-8503, Japan*
5) *Department of Geological Sciences, Stanford University, California 94305, USA*
6) *Stanford Institute for Materials and Energy Sciences, SLAC National Accelerator Laboratory, 2575 Sand Hill Road, Menlo Park, California 94025, USA*
7) *Research Reactor Institute, Kyoto University, Kumatori-cho, Sennan-gun,Osaka 590-0494, Japan*

*Present address: IoT Devices Research Laboratories, NEC Corporation, Tsukuba, Ibaraki 305-8501, Japan
**Present address: Department of Materials Science and Engineering, Kitami Institute of Technology, Kitami, Hokkaido 090-8507, Japan

**Corresponding Author**: Yoichi Kamihara

Address: Department of Applied Physics and Physico-Informatics, Faculty of Science and Technology, Keio University, 3-14-1 Hiyoshi, Yokohama 223-8522, Japan

E-mail: kamihara_yoichi@keio.jp Tel: +81-(0)45-563-1151 (dial-in:47522) Fax: +81-(0)45-566-1587

## Abstract

We unveil magnetic phases and superconducting transition temperatures ($T_c$) in an iron-based superconductor with a thick-blocking layer of a perovskite-related transition metal oxide, $Sr_2VFeAsO_{3-\delta}$ (21113V). 21113V exhibits a superconducting phase in $0.031 \leq \delta \leq 0.145$ at temperatures ($T$) < 37.1 K. Antiferromagnetic (AFM) iron sublattice are observed in $0.267 \leq \delta \leq 0.664$. A mixed valent vanadium exhibits a dominant AFM phase in $0.031 \leq \delta \leq 0.088$, while a partial ferrimagnetic (Ferri.) phase of the vanadium appears in $0.124 \leq \delta \leq 0.664$. The partial Ferri. phase becomes the most dominant for $\delta \sim 0.267$, in which the Fe shows AFM phase at $T$ < 20 K. A volume fraction of the superconducting phase is suppressed by increasing of spontaneous magnetic moments due to the partial Ferri. vanadium; i.e the magnetic phase of the vanadium dominates superconductivity in 21113V. The $T_c$–$\delta$ curve shows two maxima. The lower maximum of $T_c$ are observed for $\delta$ = 0.073. It is noted that the highest $T_c$ of 21113V appears for $\delta$ = 0.145, which exists in a phase boundary between AFM and the partial Ferri. phases of the vanadium. 21113V is a platform to verify new mechanism for enhancing $T_c$ in iron-based superconductors.

**INTRODUCTION**

The discovery of high-temperature iron-based superconductors in Mixed Anion Layered Compounds (MALC) (1-5) has triggered the search for superconducting materials. Much attention has been devoted to the newly developed superconductors with Fe-square lattices. In 2009, a family of superconducting layered iron pnictides was reported; $Ae_2M\text{Fe}Pn\text{O}_3$ (the so-called 21113 systems) with a perovskite-type layered local structure of $Ae_2M\text{O}_3$, where $Ae$ denotes an alkaline–earth metal, $M$ denotes Sc, Ti, Cr, V or another transition metals, and $Pn$ denotes P or As. The onset temperatures of the superconducting transition ($T_c^{\text{onset}}$) of the nominally synthesized $\text{Sr}_2\text{ScFePO}_3$ and $\text{Sr}_2\text{VFeAsO}_3$ are 17 K (6) and 37.2 K (7), respectively. On the other hand, $\text{Sr}_2\text{ScFeAsO}_3$ (8-11), $\text{Sr}_2\text{CrFeAsO}_3$ (8,12), $\text{Ba}_2\text{ScFeAsO}_3$ (12), and the Ca-Sc-Fe-As-O system (10) have been reported to not exhibit a superconducting phase; however, when $\text{Sc}^{3+}$ or $\text{Cr}^{3+}$ was substituted with $\text{Ti}^{4+}$, these compounds exhibit superconductivity with $T_c^{\text{onset}} = 45$ K (10), 29.2 K (13), and 37 K (10). Other superconducting materials have been reported for the 21113 systems; i.e. $\text{Sr}_2(\text{Mg, Ti})\text{FeAsO}_3$ with $T_c^{\text{onset}} = 33\text{-}39$ K (14), $\text{Ca}_2\text{AlFePO}_{3-y}$ with $T_c^{\text{onset}} = 17.1$ K, and $\text{Ca}_2\text{AlFeAsO}_{3-y}$ with $T_c^{\text{onset}} = 28.3$ K. (15) Furthermore, $\text{Sr}_3\text{Sc}_2\text{Fe}_2\text{As}_2\text{O}_5$ (16, 17) (a member of the so-called 32225 systems) and the homologous series of $\text{Ca}_{n+1}(\text{Sc, Ti})_n\text{Fe}_2\text{As}_2\text{O}_y$ ($n = 3, 4, 5$; $y \sim 3n-1$) (18) and Ca$_{n+2}$(Al, Ti)$_n$Fe$_2$As$_2$O$_y$ ($n$ = 2, 3, 4; $y$ = 3$n$) (19) were reported as compounds with thicker perovskite-related blocking layers compared to those of the 21113 systems.

Among these compounds, Sr$_2$VFeAsO$_{3-\delta}$ has attracted attention as a practical superconducting material because of its large upper critical magnetic field ($\mu_0H_{c2}$) (7,20). The large $\mu_0H_{c2}$ is suitable for applications involving high magnetic fields. As shown in Fig. 1a, the crystal structure of Sr$_2$VFeAsO$_{3-\delta}$ is described as a tetragonal lattice with FeAs carrier conducting layers sandwiched by Sr$_2$VO$_{3-\delta}$ perovskite-related carrier-blocking layers. Sr$_2$VFeAsO$_{3-\delta}$, which has been reported with nominal chemical compositions, shows superconducting transitons at $T_c^{onset} \leq$ 37.2 K (7,20-26) under ambient pressure and at $T_c^{onset}$ = 46.0 K (21) under high pressure. Munevar *et al.* and Hummel *et al.* reported that the superconducting Sr$_2$VFeAsO$_{3-\delta}$ shows a quadrupole doublet for the $^{57}$Fe Mössbauer spectra at 1.5–300 K (26,27), indicating no spontaneous magnetic moment at the Fe sublattice. These reports are in contrast with normal conducting Sr$_2$ScFeAsO$_3$ (26) and Sr$_2$CrFeAsO$_3$ (28). The $^{57}$Fe Mössbauer spectrum of Sr$_2$ScFeAsO$_3$ (26) exhibits sextet lines, indicating an existence of a finite internal magnetic field due to magnetic moments of the Fe sublattice. Furthermore, Tegel *et al.* reported that $^{57}$Fe Mössbauer spectrum of Sr$_2$CrFeAsO$_3$ (28) shows substantial broadening at 4.2 K, which could be fitted to various internal magnetic fields. The $^{57}$Fe

Mössbauer spectra of $Sr_2M FeAsO_3$ ($M$ = Sc, Cr) suggest that a magnetic phase of the Fe sublattice in $Sr_2VFeAsO_{3-\delta}$ are still controversial (20, 22-24, 26, 27, 29). Therefore, constructing magnetic and electronic phase diagrams of $Sr_2VFeAsO_{3-\delta}$ requires systematic study of its chemical compositions and element-specific magnetic measurements.

In this study, polycrystalline $Sr_2VFeAsO_{3-\delta}$ samples were prepared with various $\delta$. The samples were characterized by X-ray diffraction (XRD), X-ray fluorescence (XRF), $^{57}$Fe Mössbauer spectroscopy (30), thermal gas desorption, resistivity ($\rho$) and magnetization measurements at various temperatures ($T$) to define superconducting and magnetic phase transition temperatures. An antiferromagnetic (AFM) phase of Fe appears in normal conducitng samples with $0.267 \leq \delta \leq 0.664$. Such a topology between the magnetic phase of Fe and superconduciting phase is similar to those of other high-$T_c$ superconductors including $Re FeAsO_{1-x}(F, H)_x$ ($Re$: rare earth) (31,32) and $YBa_2Cu_3O_{7-\delta}$ (33-35), whereas a dominant and/or partial AFM phase of the V appears in samples with $0.031 \leq \delta \leq 0.267$, and a partial and/or dominant ferrimagnetic (Ferri.) phase of the V appears in samples with $0.124 \leq \delta \leq 0.664$. Density functional theory (DFT) calculations for the magnetic phase of $Sr_2VFeAsO_{3-\delta}$ support our interpretation. (23)

$T_c$ as a function of $\delta$ shows two maxima. The lower maximum coexists with the AFM

phase of the V sublattice. The higher maximum appears near a boundary between the AFM and the Ferri. phases of the V sublattice. These results clarify a topology between the superconductivity and the magnetism in MALC with perovskite-related magnetic oxide layers.

**Experimental**

Polycrystalline samples were prepared via a two-step solid-state reaction in a sealed silica tube using dehydrated SrO, FeAs, $V_2O_5$, and V as starting materials. The dehydrated SrO was prepared by heating commercial $Sr(OH)_2 \cdot 8H_2O$ powder (Sigma Aldrich Japan Co. Ltd; 99.995 wt.%) at 900°C for 10 h in air. To obtain FeAs powder, Fe (Ko-jundo Chemical Laboratory; ≥ 99.99 wt.%) and As (Ko-jundo Chemical Laboratory; 99.9999 wt.%) were mixed in a stoichiometric ratio of 1 : 1 and heated at 600°C for 10 h. in an evacuated silica tube. They were then mixed with dehydrated SrO, $V_2O_5$ (Sigma Aldrich Japan Co. Ltd; 99.99 wt.%), and V (Sigma Aldrich Japan Co. Ltd; 99.9 wt.%) in the formula $Sr_2VFeAsO_{3-\delta}$, and the powder was pressed into a pellet. These procedures were carried out in an Ar-filled glove box (MIWA Mfg; $O_2$, $H_2O \leq 1$ ppm). The pellet was loaded into an alumina boat and sealed in an evacuated silica tube. The pellet was then heated to 1050–1300°C at a rate of 30°Ch$^{-1}$, maintained at the target temperature for 20 h. and then slowly cooled to room temperature (RT). The surface of the samples was polished using abrasive paper sheets coated with SiC (Sankyo Rikagaku Co. Ltd; grit nos.400 and 1000). The phase purity and lattice constants of the resulting powders were examined by powder X-ray diffraction (XRD) (Rigaku; RINT2500Ultra18) using Cu Kα radiation from a rotating anode; the samples were scanned at RT, and the structures were refined via the Rietveld method using the RIETAN-FP-Venus program (36). In the case of $\delta = 0.509$ sample, Rietveld analysis reveals that the total amount of impurity phases is ~7 vol. %. Our analysis procedures are summarized in the supplementary information section1. See Figs S1-S3.

 The low temperature XRD (LTXRD) measurements were performed with synchrotron radiation at beamline 16-BMD of the Advanced Photon Source (APS) in Argonne National Laboratory (ANL); the measurement were carried out at $5 \leq T \leq 30$ K and at 290 K using diamond anvil cell in a cryostat.

The oxygen contents and valence states of vanadium ions in representative $Sr_2VFeAsO_{3-\delta}$ samples were examined by X-ray fluorescence (XRF) using wavelength-dispersive X-ray spectrometry (WDX; Rigaku ZSX100e) and thermal gas desorption spectrometry. Our analysis procedures are summarized in the supplementary information section1. See Fig. S4.

DC electrical resistivity ($\rho$) was measured by a four-probe technique with Au electrodes. The $\rho$ values of several superconducting samples were measured under magnetic fluxes ($\mu_0 H$ = 0–9 T) at $2 \leq T \leq 300$ K using a Quantum Design physical property measurement system (PPMS). Specific heat measurements were carried out at $2 \leq T \leq 300$ K using a relaxation method on the PPMS. Magnetization measurements were performed on a superconducting quantum interference device (SQUID) magnetometer (Quantum Design magnetic property measurement system (MPMS)) at $1.8 \leq T \leq 400$ K from −5.5 to 5.5 T. $^{57}$Fe Mössbauer spectra (MS) were obtained using conventional equipment with $^{57}$Co source from 2.5 K to 300 K. (30) The $^{57}$Fe MS were measured for $\delta$ = 0.124, 0.232, 0.237, 0.267, 0.509, and 0.631 samples. IS values of each sample are determined relative to that of $\alpha$-Fe. Details of our analysis are summarized in supplementary information section 2. See Fig. S6,7.

The magnetic properties of $Sr_2VFeAsO_{3-\delta}$ were evaluated in conjunction with DFT calculations within the generalized gradient approximation (PBE (37)), using code the Vienna Ab initio Simulation Package (VASP) (38, 39). As shown in Fig. 1c–f, the unit cell was extended to a 32-atom $\sqrt{2}\times\sqrt{2}\times1$ supercell to describe the checkerboard-type antiferromagnetism in the vanadium layer, which was reported by Nakamura *et al*. (40). The core valence interaction was treated within the projector-augmented wave scheme (41). A plane-wave cut-off energy of 500 eV was used, and the Brillouin zone was sampled by a 3 × 3 × 1 Monkhorst-Pack grid with $\sigma = 0.05$ eV. To account for the correlated 3d orbitals of vanadium, calculations were performed at the DFT+$U$ (42) level. We chose $U = 5.5$ eV and $J = 0.93$ eV, following Nakamura *et al.* (40). Experimental lattice constants were used for the calculations.

Oxygen-deficient structures with $\delta = 0.25$ and 0.50 were modeled by removing one or two 1-coordinated oxygen atoms from the supercell, as suggested by Suetin *et al.*(43). The internal coordinates were relaxed until Hellmann-Feynman forces were reduced to less than 0.02 eV/Å. The initial magnetic moments on the V and the Fe were varied to evaluate formation enthalpy in several magnetic configurations.

**RESULTS**

Figure 2 shows the powder XRD patterns of $Sr_2VFeAsO_{3-\delta}$ ($\delta = 0.124, 0.237,$ and $0.509$). Almost diffraction peaks are assigned to Bragg diffraction angles associated with the tetragonal $Sr_2VFeAsO_{3-\delta}$ phase, although several weak peaks that can be attributed to $SrV_2O_6$, $Sr_2VO_4$, FeAs, and $Fe_3O_4$ (indicated by black, blue, green, and wine-red arrows) are also observed. The polycrystalline samples are predominantly composed of the tetragonal $Sr_2VFeAsO_{3-\delta}$ phase, while the chemical composition of the primary phase are different from the nominal composition due to the off-stoichiometry of oxygen and the appearance of the second phases. In this study, values of the $\delta$ are determined using lattice volume ($V$), provided a linear relation between $\delta$ and $V$. Figure 3 shows lattice constants ($a$, $c$) and $V$ as a function of $\delta$ at room temperatures (RT). $a$ and $c$ systematically change with $\delta$. $a$ decreases with increasing $\delta$, whereas $c$ increases. The $a$–$\delta$ and $c$–$\delta$ curves exhibit a distinct hump and a dip at $\delta$ ~0.07, ~0.14, ~0.23, and ~0.62. Single-phase stoichiometric polycrystalline $Sr_2VFeAsO_3$ has not been obtained.

Figure 4a shows the V-K$\alpha_{1,2}$ XRF spectra of the $\delta = 0.267$, 0.664 samples, pure V, and the vanadium oxides ($V_2O_3$, $V_2O_5$). Each spectrum was fitted by least-squares method. The energy of the V-K$\alpha_1$ for $Sr_2VFeAsO_{3-\delta}$, $V_2O_3$, and $V_2O_5$ slightly shifts to lower energies than that of the pure V. V-K$\alpha_1$ of the pure V is denoted by the vertical solid line at 4951.170(3) eV. Therefore, the chemical shifts of V ions in $Sr_2VFeAsO_{3-\delta}$, $V_2O_3$ and $V_2O_5$ were determined as the difference of the energy between them and the pure V. As shown in Fig. 4b, provided that nominal valences of the pure V, $V_2O_3$, and $V_2O_5$ were 0, 3, and 5 respectively, those of the $\delta = 0.267$, 0.664 samples were determined as 2.6(1) and 2.0(1).

Figure 5a shows the temperature ($T$) dependences of the electrical resistivity ($\rho$) for $Sr_2VFeAsO_{3-\delta}$ samples at $2 \leq T \leq 300$ K. As shown in Figs. 5a and 5b, the $0.204 \leq \delta \leq$ 0.664 samples show normal conductivity at $T \geq 2$ K, whereas the $0.204 \leq \delta \leq 0.237$ samples exhibit sharp decreases of $\rho$, which are verified as superconducting trace (SC-trace); i.e. $\delta = 0.204$-$0.237$ samples show a chemically inhomogeneous and an phase-segregated electronic state, and contain both major normal conducting grains and minor superconducting grains. For these normal conducting samples, $\partial \rho/\partial T$ values of the $0.204 \leq \delta \leq 0.237$ samples are positive at $20 \leq T \leq 300$ K, whereas the $\partial \rho/\partial T$ of the $0.509 \leq \delta \leq 0.664$ samples exhibit a semiconductor-like negative values at $T > $ ~10 K. $\rho$-$T$ curves of the $0.267 \leq \delta \leq 0.664$ samples show anomalous maxima at $T$~10 K, as denoted by downward open triangles ($T_{max}$). The order of $\rho$ of the $0.267 \leq \delta \leq 0.664$ samples does not change at $T < T_{max}$. Such a decrease of $\rho$ at $T_{max}$ is possibly due to the local crystallographic phase transition.

$\rho$-$T$ curves of the $0.031 \leq \delta \leq 0.145$ samples exhibit superconducting transitions with $30 \leq T_c^{onset} \leq 37$ K, as denoted by the closed downward triangles. The bulk $T_c$ can be defined as $T_c^{mid}$ of these samples; See Table S2 in the supplementary information. The $\rho$–$T$ curves of the $0.031 \leq \delta \leq 0.267$ samples exhibit anomalous kink at ~200 K, denoted by upward green triangles defined as $T_{anom}$. The $T_{anom}$ of the $0.204 \leq \delta \leq 0.267$ samples with the SC-trace is less clear than those of the superconducting $0.031 \leq \delta \leq 0.145$ samples. $T_{anom}$ are not observed for the normal conducting $0.509 \leq \delta \leq 0.664$ samples without a SC-trace. $T_{anom}$ are clearly correlated with the appearance of superconducting phases.

Figure 6a shows temperature dependence of spontaneous magnetization ($M_S$) of the $\delta$ = 0.124, 0.204, 0.232, 0.237, 0.267, 0.509, 0.631, and 0.664 samples at $1.8 \leq T \leq 400$ K. In the spontaneous magnetization ($M_S$) vs. $T$ curves, the magnetic phase transition temperatures, which are determined as the intersections where the $M_S$ is equal to 0.01 $\mu_B$ per formula unit (f.u.), were defined as $T_{ferri}$. The $M_S$ of the $0.124 \leq \delta \leq 0.664$ samples reach 0.011–0.53 $\mu_B$ (f.u.)$^{-1}$ at 1.8 K. As shown in Fig. 6b, the largest values of the $M_S$ and $T_{ferri}$ were observed for the normal conducting sample with $\delta$ = 0.267. The $M_S$ value of the $\delta$ = 0.267 sample is sufficiently large to verify intrinsic magnetism, whereas the intrinsic $M_S \leq 0.01$ $\mu_B$ (f.u.)$^{-1}$ can not be distingusished from extrinsic magnetic moments due to ferromagnetic second phases. $T_{ferri}$ shows several maxima in the $T_{ferri}$-$\delta$ curve.

As shown in Fig. 6c, the $\delta$ = 0.088, 0.124, and 0.145 samples exhibit a superconducting volume fraction greater than 10 vol.% at 1.8 K, whereas the normal conducting $0.204 \leq \delta \leq 0.237$ samples show a small superconducinting volume fraction less than 10 vol.%. A diamagnetic susceptibility was not observed for the $0.267 \leq \delta \leq 0.664$ samples. Thus, $Sr_2VFeAsO_{3-\delta}$ exhibits bulk superconductivity for the $0.031 \leq \delta \leq$ 0.145 samples and a normal conductivity for the $0.204 \leq \delta \leq 0.664$ samples.

Figure 7 shows the temperature dependence of the $^{57}$Fe Mössbauer spectra for the $\delta = 0.232, 0.267$, and 0.509 samples as representative. The $^{57}$Fe Mössbauer spectra of other samples are summarized in the supplementary information (Fig. S8). The $^{57}$Fe Mössbauer spectra are composed of quadrupole doublet absorption lines at 300 K for each sample, indicating the paramagnetic (PM) phase of the Fe sublattice. Asymmetry doublet absorption lines of the samples are derived from the anisotropy of crystal orientation (44) for $Sr_2VFeAsO_{3-\delta}$. $^{57}$Fe Mössbauer spectra of the $\delta = 0.232$ sample were fitted to a doublet pattern at $2.5 \leq T \leq 300$ K without an internal magnetic field ($B_{int}$), indicating the PM Fe in the $\delta = 0.232$ sample. The $\delta = 0.124, 0.232$, and 0.237 samples also exhibit the PM Fe at $T \geq 4.2$ K. (See Fig. S6) The $^{57}$Fe Mössbauer spectra of the $\delta = 0.267$ and 0.509 samples were fitted to a doublet pattern from 30 to 300 K. The absorption lines became broader at $T\sim10$ K, indicating that spontaneous magnetic moments appear in the Fe-sublattice. The broad absorption lines are clearly different from a sextet split spectrum showing the antiferromagnetically ordered Fe sublattice which is typically observed for mother compounds of iron-based superconductors; LaFeAsO (30) and $BaFe_2As_2$ (45). As shown in Fig. 8a, the full-width at half-maximum (FWHM) of the $\delta = 0.267, 0.509, 0.631$ samples substantially increase at $T < 20$ K. The increasing FWHM indicates finite distribution of $B_{int}$ in these samples. The broad absorption lines also indicate that both of magnetic Fe and nonmagnetic Fe appear in a same chemical composition. Given the wide distribution of $B_{int}$, the broad absorption lines can be fitted sufficiently well enough to quantify mean values of the $B_{int}$. In the present study, Hesse and Rübertsch's method (46) was applied to obtain the distribution of $B_{int}$. Figure 8a shows temperature dependence of the isomer shift (IS), quadrupole splitting (QS), and FWHM for the $\delta$ = 0.124, 0.232, 0.237, 0.267, 0.509, and 0.631 samples. As shown in Fig. 8a, the IS value of the superconducting $\delta$ = 0.124 sample is the smaller than those of the normal conducting $\delta$ = 0.232, 0.237, 0.267, 0.509 and 0.631 samples at RT. The smaller IS value of the $\delta$ = 0.124 sample indicates higher s-electron densities around the nucleus compared to the other samples. In terms of the Debye approximation of the lattice vibrations, Debye temperatures were obtained at 454(29) K for the $\delta$ = 0.232 sample and 362(12) K for the $\delta$ = 0.509 sample from IS–$T$ curves. The increase of $\delta$ correlates with elastic properties of $Sr_2VFeAsO_{3-\delta}$. The upper inset of Fig. 8a indicates that the IS-$T$ curve of $\delta$ = 0.509 sample shows an anomalous decreasing at $T$ < 20 K with decreasing $T$. In general, IS for a sample, which does not show a phase transition, increases and saturates at low temperatures alike theoretical curves drawn in the upper inset of Fig. 8a. The anomalous decreasing IS indicates that there is a phase transition at $T < 20$ K for $\delta = 0.509$ sample. Indeed, several samples also show a kink in QS-$T$ curves and a sudden increasing of FWHM-$T$ curves, indicating an appearance of anisotropic electric field gradients and AFM phases of the Fe sublattice. Figure 8b shows the temperature ($T$) dependences of the square root of the mean squared amplitude of the internal magnetic field ($\sqrt{<B_{int}^2>}$) due to an AFM phase of the Fe in the $\delta = 0.267, 0.509,$ and $0.631$. $\sqrt{<B_{int}^2>}$ behaves as a function of $T$ according an empirical formula (47). The Néel temperatures ($T_N$) of the Fe are found to be $T_N = 11.9$ K, $17.5$ K and $23.1$ K with $\sqrt{<B_{int}^2>} = 4.6$ T, $7.3$ T and $7.5$ T correspoinding to average magnetic moments of $0.31$, $0.48$ and $0.5$ $\mu_B$ Fe$^{-1}$ at 0 K in $\delta = 0.267, 0.509$ and $0.631$ samples, respectively. These average magnetic moments were obtained by a conversion factor CF = 15 T$\mu_B^{-1}$ (48). The insets of Fig. 8b demonstrate a wide-range distribution of the $B_{int}$ of Fe in these samples (49, 50). The increasing $\delta$ enhances $T_N$ of Fe in Sr$_2$VFeAsO$_{3-\delta}$.

The normal conducting $\delta = 0.267, 0.509,$ and $0.631$ samples show AFM phases of the Fe sublattice at $T < \sim 20$ K. The normal conducting $\delta = 0.267, 0.509,$ and $0.631$ samples also exhibit $M$s at $T < T_{ferri} \approx 308, 297,$ and $158$ K. Such a discrepancy between the $T_N$ of Fe and both of $T_{ferri}$ indicates that the $T_{ferri}$ are almost irrelevant to the magnetic phase of the Fe sublattice. The $T_{ferri}$ is ascribed to a magnetic phase of the V sublattice. $T_{anom}$ are also irrelevant to magnetic phase of the Fe and likely correspond to the magnetic transition temperatures of the V sublattice in $Sr_2VFeAsO_{3-\delta}$, $T_N$ and/or $T_{ferri.}$ of V.

Figures 9a and 9b show temperature dependences of the molar heat capacity ($C_{mol}$), and excess contribution of heat capacity ($C_{ex}$) that is obtained by subtracting both Debye's phonon and Sommerfeld's normal-conducting contributions from $C_{mol}$. Figure 9c shows the differential length ($\Delta L$) of the lattice constants ($a$, $c$) for the normal conducting $\delta = 0.509$ sample. The appearance of the $C_{ex}$ are due to a transition of electronic and magnetic phase that has a predominant contribution to the thermal properties of solids at low temperatures. The $C_{ex}$ shows a maximum at $T \approx 10$ K and decreases with increasing $T$ and converges to zero at $T \approx 20$ K. The $C_{ex}$–$T$ curve correlates with the magnetic phase transition at $T_N$ of Fe in $\delta = 0.509$ sample. As shown in Fig. 9c, the temperature dependence of lattice constants $a$ and $c$ also show an anomalous kink at ~10 K. The anomalous kink is understood as a crystallographic local structure transition in $Sr_2VFeAsO_{3-\delta}$. These results indicate that the anomalous kink of ($a$, $c$)-$T$ curves clearly correlates with $T_{max}$ in the $\rho$-$T$ curve, and a formation enthalpy of the lattice in $Sr_2VFeAsO_{3-\delta}$. Various temperatures, where $T_N$ ($\approx T_{anom}$), $T_{ferri}$ of V, $T_N$ of Fe, $T_{max}$, and $T_c$ are plotted against $\delta$ in Fig. 10. The finite $\delta$ was inevitable for our samples. Figure 10a shows magnetic phase diagram of the V in $Sr_2VFeAsO_{3-\delta}$. The V sublattice shows an AFM phase at $\delta = 0.031$-$0.267$ region. The AFM phase of the V are verified by both of a report by Tatematsu et al (23). As shown in Fig. 7 and Fig. S6, A PM phase of the Fe sublattice is verified at 20-300 K for the samples in the present work. $\delta = 0.124, 0.204, 0.232, 0.237, 0.267, 0.509, 0.631$, and $0.664$ samples show finite $M_S$ at $T < T_{ferri}$. The $T_{ferri}$ shows maximum at $T = 304$ K at $\delta = 0.267$. The $M_S$ decreases with increasing $\delta$ at $\delta > 0.267$. It is noted that the AFM and Ferri. phases of the V are observed for $\delta = 0.124$–$0.267$ indicating a phase segregated magnetic state of the V sublattice. Such a inhomogeneous state should be discussed as an intrinsic property of off-stoichiometric compounds. Indeed, an apparent valence of the vanadium is mixed valence that is an essential electronic state for ferrimagnetic phase of the vanadium in $Sr_2VFeAsO_{3-\delta}$.

The optimum oxygen deficiencies that give maximum $T_c$s exist at $\delta = 0.073$ and at $\delta = 0.145$. $Sr_2VFeAsO_{3-\delta}$ exhibits a bulk superconducting (SC) phase below the $T_c^{mid}$–$\delta$ curve for the $0.031 \leq \delta \leq 0.145$ samples. A normal conducting phase is observed for the $0.204 \leq \delta \leq 0.664$ samples, although SC-traces appear in the $0.204 \leq \delta \leq 0.237$ samples. The electrical phase segregation, which is mainly due to the inhomogenous chemical state, is consistent with the inhomogeneous magnetic states of the vanadium in $Sr_2VFeAsO_{3-\delta}$ with $\delta = 0.124$–$0.267$. The PM phase of the Fe sublattice is observed for $Sr_2VFeAsO_{3-\delta}$ with $0.031 \leq \delta \leq 0.237$ at 2 K, while the normal conducting $\delta = 0.267$, 0.509, and 0.631 samples exhibit AFM phases of the Fe at $T < \sim 20$ K, indicating that SC phase does not coexist with AFM phase of the Fe.

**DISCUSSION**

As shonw in Fig. 3, an expansion of the lattice volume with increasing $\delta$, which is observed for $Sr_2VFeAsO_{3-\delta}$, have also been observed for perovskite-type and perovskite-related compounds such as $SrTiO_{3-\delta}$ (51), $YBa_2Cu_3O_{7-\delta}$ (33), and $Bi_{2.1}Sr_{1.9}Ca_2Cu_3O_{10+\delta}$ (52). The lattice expansion is probably general crystallographic properties of perovskite-related compounds. An average of the bond valence sum (BVS) (53) for $V^{III}$, $V^{IV}$, and $V^{V}$ in $\delta = 0.509$ sample is 2.81, which is smaller than the average BVS reported by Cao *et al.*(24) The value of the As–Fe–As bond angle ($\alpha$) in the FeAs layer (see Fig. 1b and Table S1 in supplementary information) is 106.43(9)° for the $\delta = 0.509$ sample. The $\alpha$ indicates that the $FeAs_4$ tetrahdron is distorted from a regular tetrahedron with $\alpha = 109.47°$. (7,24)

Figure 10 indicates that superconductivity in $Sr_2VFeAsO_{3-\delta}$ is induced by optimum $\delta$ accompanied with suppression of the magnetic phase of the Fe. Indeed, Nakamura *et al.* proposed theoretical Fermi surface nesting (40), that suggests the emergence of a AFM phase of Fe. Such a relation between the AFM phase and the SC phase of the Fe is similar to that of the "1111" compounds (2,48). The AFM phase of the Fe may be understood as a spin density wave.

DFT calculations quantify magnetic ground states in the vanadium and the Fe sublattices based on virtually defined supercells of $Sr_2VFeAsO_{3-\delta}$ ($\delta = 0, 0.25, 0.50$) (Figs. 1d–f).

Table 1 shows four different magnetic configurations and their total energies for $\delta = 0$. Internal coordinates are fixed. The differences of the formation enthalpy ($\Delta E$) among the four configurations are due to differences in their charge density distributions. We define the checkerboard-AFM phase of the V as c-AF on V. The c-AF on V indicates that the neighboring magnetic V ions on the same layer always have opposite moments. An AFM ordering between V layers and neighboring V layers having opposite moments is defined as A-AF on V. Further we define the stripe-AFM phase as s-AF on Fe. The s-AF indicates that magnetic ions on the same layer align line by line, as proposed by Nakamura *et al* (40). The $\Delta E$ is referenced to the most stable structure, c-AF on V and s-AF on Fe, in accord with Nakamura's results (40). The variation in $\Delta E$ represents that several stable solutions are calculated even for the same magnetic phase. They exhibit several $\Delta E$ values for a magnetic phase. The lowest energy of (A-AF on V, PM on Fe) and (c-AF on V, PM on Fe) are 175 and 183 meV; i.e. PM on Fe is not a ground state.
s-AF is the most stable for the Fe. The lowest energy of (c-AF on V, s-AF on Fe) is
lower than those of (A-AF on V, s-AF on Fe) by 4 meV. On the basis of this
comparison, we conclude that the stability of A-AF on V and c-AF on V can be
switched by slight change of chemical composition, because of these quite close
energies, irrespective of a magnetic phase of the Fe. Our results partially supports the
work of Nakamura *et al*. (40). The subtle differences between our results and Nakamura
et al`s results are caused by different exchange correlation potentials and/or initial
conditions; e.g. values of magnetic moments adopted on V and Fe. Clarification of this
issue requires further computational analysis on a larger magnetic supercell of
$Sr_2VFeAsO_{3-\delta}$. In a consequence of the analysis, we define a ferromagnetic (FM) phase
and a Ferri. phase of the V in oxygen deficient supercells. Table 2 shows several
different magnetic configurations of the V and their $\Delta E$ for $\delta = 0, 0.25, 0.50$, provided
that magnetic configurations of the Fe is fixed as s-AF. In this case, we relaxed the
internal coordinates. The noticeable relaxation occurred only along the *c*-axis within
0.02 nm, and the relaxation trends were almost the same among each configuration. As
shown in the first block of Table 2, the $\Delta E$ of several magnetic configurations on the V
layer show close values. The theoretical magnetic phases of the V exhibit similar $\Delta E$ in A-AF, FM, c-AFs, and Ferri. phases. For oxygen-deficient conditions $\delta = 0.25$ and 0.50, a clear difference is observed in the magnitude of the magnetic moment of the V, although the similarity of $\Delta E$ remains the same. Upon removal of the oxygen, the V loses one of its bonded O in the supercell. Both of A-AF phase of the V for $\delta = 0.25$ and c-AF phase of the V for $\delta = 0.50$ behave as a kind of Ferri. phase. On the basis of these results, we argue that $d$ electron is rather localized on around the V ions exhibiting magnetic moment from 1.9 $\mu_B$ to 2.3 $\mu_B$. Such different magnetic moments of the V breaks the magnetic balance of the V sublattice, possibly stabilizing Ferri. phase of V. Our calculations verify that the magnetic ground states of the V are A-AF for $\delta = 0$, Ferri. for $\delta = 0.25$, and c-AF with small spontaneous magnetic moments for $\delta = 0.50$ in the virtual superlattice of $Sr_2VFeAsO_{3-\delta}$. The calculated stabilities of magnetic phases of the V are consistent with experimental results for $\delta \sim 0$, $\sim 0.25$, and $\sim 0.50$. As shown in Figs 6, 10, and Tables, the largest spontaneous magnetic moments is observed for $\delta = 0.267$ sample.

The $T_{max}$–$\delta$ curve is demonstrated in Fig. 10. Here we assume $T_{max}$ to be equal to a temperature of a local crystallographic transition, probed as the anomalous kinks in ($a$, $c$)–$T$ curves for $\delta = 0.509$. As shown in Figs 5a, 9b, and 9c, the $T_{max}$ and $C_{ex}$ are likely due to the local crystallographic phase transition at 10 K for $\delta = 0.509$. The possible local crystallographic transition is consistent with results obtained by detailed analysis on $^{57}$Fe Mössbauer spectra describted in supplementary information section 3.

**CONCLUSION**

Here in, we prepared well-characterized polycrystalline $Sr_2VFeAsO_{3-\delta}$ to measure electrical resistivity, magnetization, and $^{57}$Fe Mössbauer spctra. Density functional theory (DFT) calculations support a presence of several magnetic phases for the V sublattice. Provided there is a linear relation between $\delta$ and the lattice volume, $a$–$\delta$ and $c$–$\delta$ curves change systematically with $\delta$. Bulk superconductivity are observed in the samples with $0.031 \leq \delta \leq 0.145$. The samples with $0.204 \leq \delta \leq 0.664$ show normal conductivity at $T > 2$ K, although SC-traces were observed in the samples with $0.204 \leq \delta \leq 0.237$. These SC-traces are mainly due to the magnetic phase segregation. The highest $T_c^{onset}$ and $T_c^{mid}$ were 37.1 K and 34.1 K, respectively, for $\delta = 0.145$ sample. A spontaneous magnetic moment ($M_S$) of the V appears for the samples with $0.124 \leq \delta \leq 0.631$. The $M_S$ exhibits a maximum value of ~0.5 $\mu_B$ (f.u.)$^{-1}$ for $\delta = 0.267$ sample. $0.124 \leq \delta \leq 0.631$ samples exhibit magnetic transition temperatures of the V ranging from 25 to 308 K. $^{57}$Fe Mössbauer spctra of $Sr_2VFeAsO_{3-\delta}$ exhibit PM phase of the Fe sublattice for superconducting $\delta = 0.124$ sample and normalconducting samples with $\delta = 0.232$ and 0.237, whereas AFM phases of Fe appear at $T < 11.9$ K, 17.5 K, and 23.1 K for the samples with $\delta = 0.267$, 0.509, and 0.631. Our DFT calculations show that a stable AFM phase of the V can be switched, depending on the $\delta$, because the stability of the magnetic phases are quite similar in energy. In addition, our DFT calculations indicate that the Ferri. phase of the V is the most stable in the samples with $\delta = 0.25$. Based on these results, an electronic and magnetic phase diagram of $Sr_2VFeAsO_{3-\delta}$ is constructed with respect to $\delta$. This phase diagram provides an evidence that there is no coexisting AFM phase of the Fe sublattice with SC, although a higher maximum of $T_c$ appears near a boundary between the AFM and Ferri. phases of V. Magnetic and valence states of the V dominate $T_c$ of $Sr_2VFeAsO_{3-\delta}$ indirectly. This phase diagram extends our knowledge to understand optimum chemical compositions for the superconductivity of iron-based MALC and should help understand the mechanism of enhancing $T_c$ in multinary transition metals-based MALC; promising materials for superconducting applications under high magnetic fields.

This article contains supplementary information online refering references 54-65 at URL:******.

**ACKNOWLEDGMENTS.** This work was supported by the research grant of Keio University, the Sumitomo foundation, and the Ministry of Education, Culture, Sports, Science, and Technology (MEXT) through the Element Stragegy Initiative to Form Core Research Center. This work has been performed by using facilities of the Research Reactor Institute, Kyoto University. We thank D. Ikuta for help with the XRD experiments. This research is partially funded by the US Department of Energy (DOE), Office of Basic Energy Sciences (BES). S.H.and W.L.M. are supported by the US Department of Energy (DOE), Office of Basic Energy Sciences (BES), Division of Materials Sciences and Engineering, under Contact No.DE-AC02-76SF00515. H. Hi. was also supported by the Japan Sociery for the Promotion of Science (JSPS) through a Grant-in-Aid for scientific Research on Innovative Areas "Nano Informatics" (Grant Number 25106007), a Grant-in-Aid for Scientific Research (A) (grant no. 17H01318) and support for Tokyotech Advanced Research (STAR). YK was also supported by the JSPS through a Grant-in-Aid for scientific Research (A) (grant no. 15H01998, 17H03239), and Core-to-Core Program "Isotope spintronics".

**Figure captions**

Fig. 1. Crystallographic structure (a), local structures, and supercell structures in $Sr_2VFeAsO_{3-\delta}$. (b) The dashed box represents a unit cell in the tetragonal phase.

Schematic molecular FeAs$_4$ in the carrier conducting FeAs layer (upper) and a VO$_4$ in the perovskite-related Sr$_2$VO$_{3-\delta}$ layer (lower). The two As–Fe–As bond angles ($\alpha$, $\beta$) and the disntance of As from the plane formed by Fe atoms ($h_{pn}$) in the FeAs layer and the V–O–V bond angle ($\gamma$) in the Sr$_2$VO$_{3-\delta}$ layer are illustrated. Atomic coordinations are summarized in the supplementary information (see Table S1). (c) Top view of the crystal structure from the *c*-axis direction. The dahed square represents a unit cell in the tetragonal phase. The solid square represents a unit cell that is extended to a supercell. The *a*-axis of the supercell is rotated by 45° from that of the tetragonal phase. The lattice constants *a* for the supercells are expanded by $\sqrt{2}$. (d)–(f) Supercell structures of Sr$_2$VFeAsO$_{3-\delta}$ ($\delta$ = 0.0, 0.25, and 0.50, respectively). In our DFT calculation, the vanadium sites are named V1, V2, V3, and V4, as shown in (d), and the blue-green colored oxygen sites show the site of oxygen defects are defined for $\delta$ = 0.25 (e) and for $\delta$ = 0.5 (f).

Fig. 2. Powder XRD patterns of representative Sr$_2$VFeAsO$_{3-\delta}$ samples ($\delta$ = 0.124, 0.237, and 0.509) at room temperature. The vertical bars at the bottom represent the calculated angles of Bragg diffraction for $\delta$ = 0.509. The black, blue, green and wine-red arrows represent the Bragg diffraction angles of the impurity phases SrV$_2$O$_6$, Sr$_2$VO$_4$, FeAs, and Fe$_3$O$_4$.

Fig. 3. Lattice constants (*a*, *c*) of Sr$_2$VFeAsO$_{3-\delta}$ samples at room temperature as functions of oxygen deficiency ($\delta$) assuming a linear relationship between *V* and $\delta$. The small black lines in the red plots show standard deviation. The $\delta$ values were determined from the lattce constants of the representative samples. (See supplementary information for XRD patterns of each sample)

Fig. 4. (a) X-ray fluorescence (XRF) V-K$\alpha$ spectra as a function of energy (*E*) for pure vanadium metal (V), Sr$_2$VFeAsO$_{3-\delta}$ ($\delta$ = 0.267 and 0.664), V$_2$O$_3$, and V$_2$O$_5$. Red circles show the observed data points, and the black, blue, and green lines denote the fitted V-K$\alpha$, V-K$\alpha_1$, and V-K$\alpha_2$ spectra, respectiviely. The vertical red line denotes the *E* value of the V-K$\alpha_1$ edge for pure V at 4951.170(3) eV as a standard *E* value. (b) Nominal valences of Sr$_2$VFeAsO$_{3-\delta}$ ($\delta$ = 0.267 and 0.664), V$_2$O$_3$, and V$_2$O$_5$ as a function of the chemical shift of V-K$\alpha_1$. The dashed line is a guide for the eyes. The inset shows an expanded view of XRF intensity as a function of *E* relative to V.

Fig. 5. (a) Electrical resistivity ($\rho$) as a function of temperature (*T*) in the range of 2−300

K for $Sr_2VFeAsO_{3-\delta}$ ($0.031 \leq \delta \leq 0.664$) samples. The $\delta$ value is indicated in the each plot. The closed downward triangles indicate the onset temperatures corresponding to the bulk superconducting transition ($T_c^{onset}$). The open downward triangles indicate the temperatures of a maximum at 10–15 K ($T_{max}$) for normal conducting sapmles. The green upward triangles indicate temperatures of an anomalous kink at ~200 K ($T_{anom}$). (b) The $\rho$-$T$ curves for the samples with $\delta$ = 0.145 and 0.204. The dashed line denotes the detection limit of our measurement.

Fig. 6. Magnetic properties of $Sr_2VFeAsO_{3-\delta}$. (a) Spontaneous magnetization ($M_S$) as a function of temperature ($T$) for the samples with $\delta$ = 0.124, 0.204, 0.232, 0.237, 0.267, 0.509, 0.631 and 0.664 (the corresponding $M_S$ values at 1.8 K are ~0.011 $\mu_B$, ~0.055 $\mu_B$, ~0.028 $\mu_B$, ~0.053 $\mu_B$, ~0.50 $\mu_B$, ~0.43 $\mu_B$, ~0.19 $\mu_B$ and ~0.22 $\mu_B$ per formula unit (f.u.)$^{-1}$, respectively). The black arrows indicate ferrimagnetic transition temperatures ($T_{ferri}$), which are defined as temperatures at $M_S$ ($T$) = 0.01 $\mu_B$ (f.u.)$^{-1}$ (indicated by the dashed line). As shown in Fig. S2, $\delta$ = 0.509 contained the magnetic impurity phase 0.19 at.% $Fe_3O_4$. The contribution of $Fe_3O_4$ was excluded from the $M_s$-$T$ curve for $\delta$ = 0.509. (b) $M_S$ at 1.8 K (closed circles) and $T_{ferri}$ (open circles) versus $\delta$. (c) Superconducting volume fraction (SVF) versus $\delta$. Inset shows molar magnetization ($M_{mol}$) versus magnetic flux density ($\mu_0H$) at 1.8 K for the samples with $\delta$ = 0.088 (red closed circles), 0.124 (purple closed squares), 0.145 (brown open triangles), 0.204 (gray open diamonds), 0.237 (green closed triangles), and 0.267 (blue open circles). The solid line indicates the magnetization of the perfect diamagnetism using magnetic susceptibility $\chi_{mol}$ = −5.81 emu for $Sr_2VFeAsO_{3-\delta}$.

Fig. 7. $^{57}Fe$ Mössbauer spectra (MS) of representative samples of $Sr_2VFeAsO_{3-\delta}$ at different temperatures indicated near the plots. (a) $\delta$ = 0.232; (b) $\delta$ = 0.267; and (c) $\delta$ = 0.509. The solid lines are fitted patterns with a wide distribution of internal magnetic field. $^{57}Fe$ MS are summarized with those of other samples including $\delta$ = 0.124 and $\delta$ = 0.631 in Fig. S7 in the supplementary information.

Fig.8. Electronic and magnetic properties of $^{57}Fe$ in $Sr_2VFeAsO_{3-\delta}$. (a) Temperature ($T$) dependences of isomer shifts (IS), quadrupole splitting (QS) and full width at half maximum (FWHM) for $Sr_2VFeAsO_{3-\delta}$ samples with $\delta$ = 0.124 (open circle), 0.232 (open square), 0.237 (open triangles), 0.267 (closed triangles), 0.509 (red closed circles), and 0.631 (closed squares). The QS and FWHM values of the samples with $\delta$ = 0.267, 0.509, and 0.631 are written only when the spectrum is a doublet. The black lines in all plots indicate standard deviations. Insets show expanded view of the Figs. See supplementary information for details of our analysis. (b) Temperature ($T$) dependence of the square root of the mean squared amplitude of internal magnetic field, $\sqrt{<B_{int}{}^2>}$, which is obtained from the distributions of the internal fields ($B_{int}$) of the samples with $\delta$ = 0.267, 0.509, and 0.631 exhibiting antiferromagnetic ordering of Fe sub-lattice at temperatures < Néel temperatures ($T_N$). The red dashed lines are fitted lines, obtained by the formula by

$$\sqrt{\langle B_{int}{}^2 \rangle} = B_0 \left(1 - \frac{T}{T_N}\right)^\alpha \text{ for } 0 \leq \frac{T}{T_N} \leq 1$$

with $\alpha$ = 0.5. Insets show histograms of $B_{int}$ distributions for each sample at 4.2 K.

Fig. 9. Temperature ($T$) dependences of molar heat capacity ($C_{mol}$) (a), excess contribution of heat capacity ($C_{ex}$) (b), and lattice length change ($\Delta L$) (c) for $Sr_2VFeAsO_{3-\delta}$ ($\delta$ = 0.509). The dashed line indicates the best fits of $C_{low} = \gamma T + \beta T^3$, where $\gamma$ and $\beta$ are the Sommerfeld coefficient and the lattice heat capacity coefficient for Debye heat capacity at low temperatures. ($\gamma$ = 65.5(3) mJ K$^{-2}$ mol$^{-1}$ and $\beta$ = 0.52(1) mJ K$^{-4}$ mol$^{-1}$). $C_{ex}$ is calculated as $C_{ex} = C_{mol} - C_{low}$. $\Delta L$ is defined as differences to lattice constants ($a$ and $c$) at 290 K.

Fig. 10. Element specific electronic and magnetic phase diagrams of $Sr_2VFeAsO_{3-\delta}$ in terms of $\delta$ and $T$. (a) Magnetic phase diagram of the V. Green closed triangles indicate ferrimagnetic (Ferri.) phase transition temperatures defined at temperatures with spontaneous magnetic moment, $M_S(T)$ = 0.01 $\mu_B$ (f.u.)$^{-1}$. The green shadowed area indicates an Ferri. phase. The blue open triangles indicate antiferromagnetic (AFM) Néel temperatures ($T_N$) assuming that the $T_N$ appears at ~$T_{anom}$ defined in Fig. 5. Indeed, Tatematsu et al reported the $T_N$ appears at around our $T_{anom}$. (23) The blue shadowed area indicates an AFM phase. The area, which is overlapped with the Ferri. and the AFM phases, denotes a region in which a chemical inhomogeneity results appearance of superconducting traces. (See Table S2 in supplymentary information) (b) Electronic and magnetic phase diagram of Fe in $Sr_2VFeAsO_{3-\delta}$. Red open circles indicate $T_c^{mid}$ for bulk superconducting samples. (See Table S2 in supplymentary information) The small black lines whithin the plots indicate standard deviation. The red shadowed area indicates superconducintg (SC) phase. The blue closed squares indicate $T_N$ determined by $^{57}$Fe Mössbauer spectroscopy. The blue shadowed area indicates an AFM phase. Black downward triangles, which are probably coupled with crystallographic local structure

transition, denote the $T_{max}$ observed for normal conducting samples.

Table 1. Magnetic moments and differences of the formation enthalpy ($\Delta E$) of several magnetic configurations for $Sr_2VFeAsO_{3-\delta}$ ($\delta = 0.0$). Internal coordinates are fixed at the reported values. (7)

Table 2. Magnetic moments and differences of the formation enthalpy ($\Delta E$) of several magnetic configurations for $Sr_2VFeAsO_{3-\delta}$ ($\delta = 0.0, 0.25,$ and $0.50$). Internal coordinates are relaxed. The Fe layer is kept at s-AF.

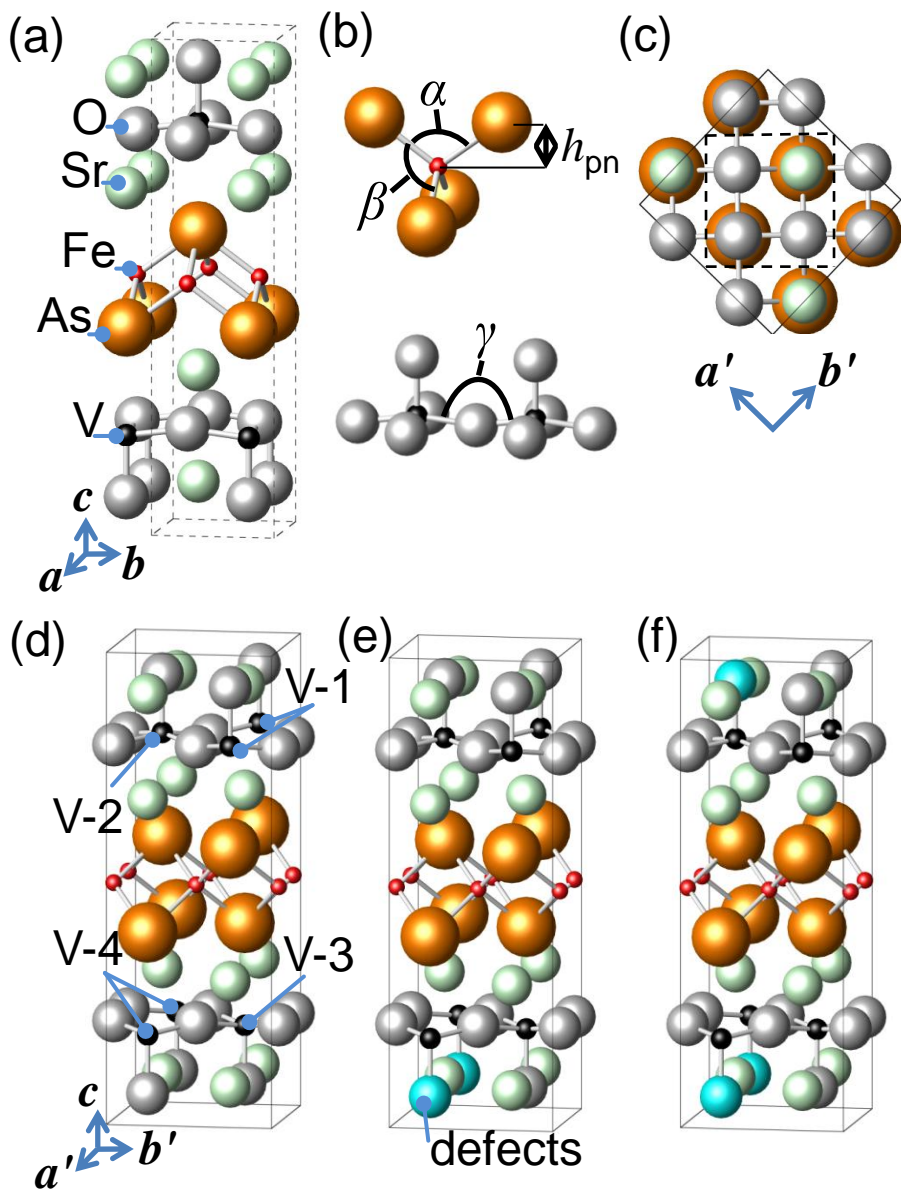

Fig. 1

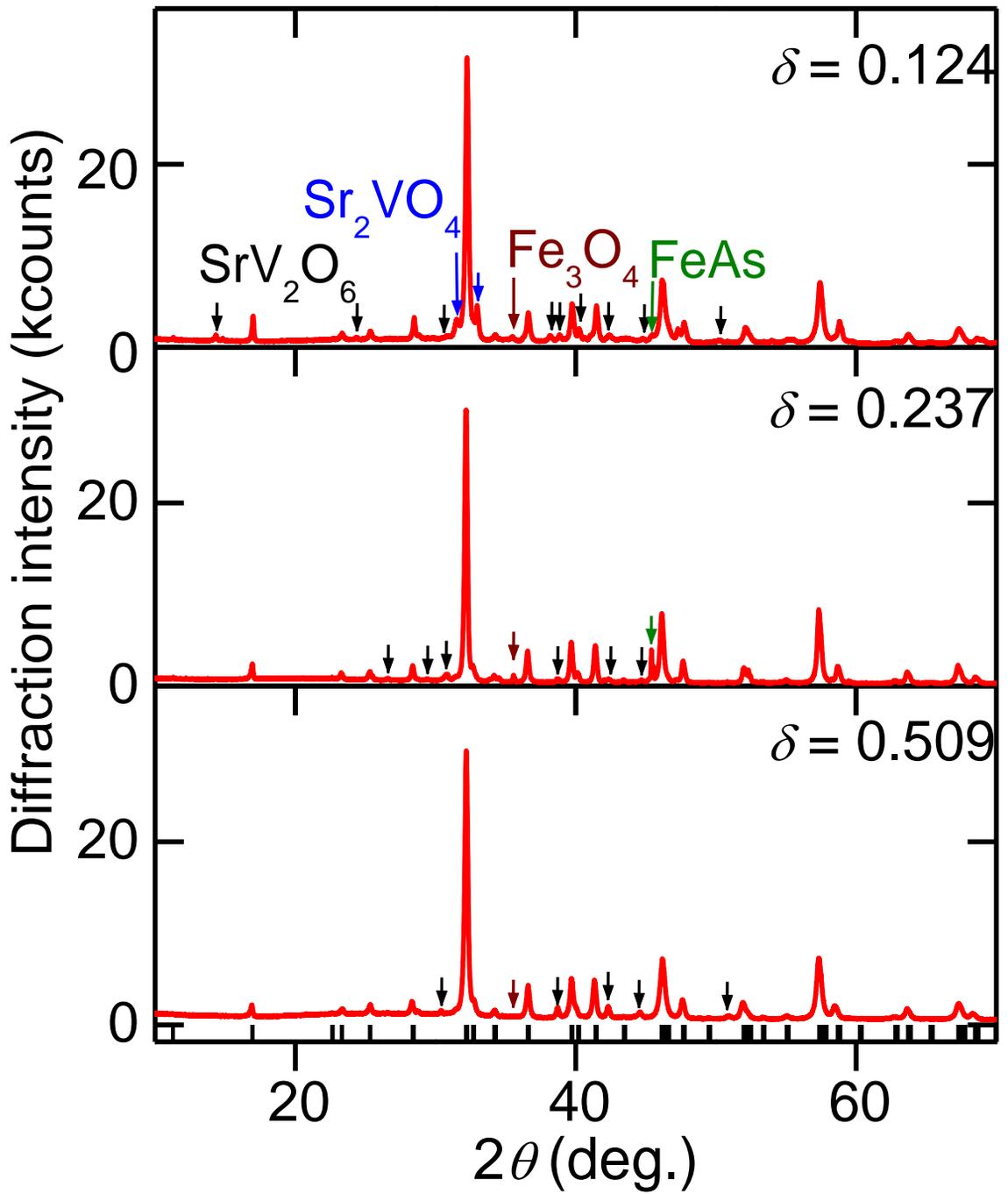

Fig. 2

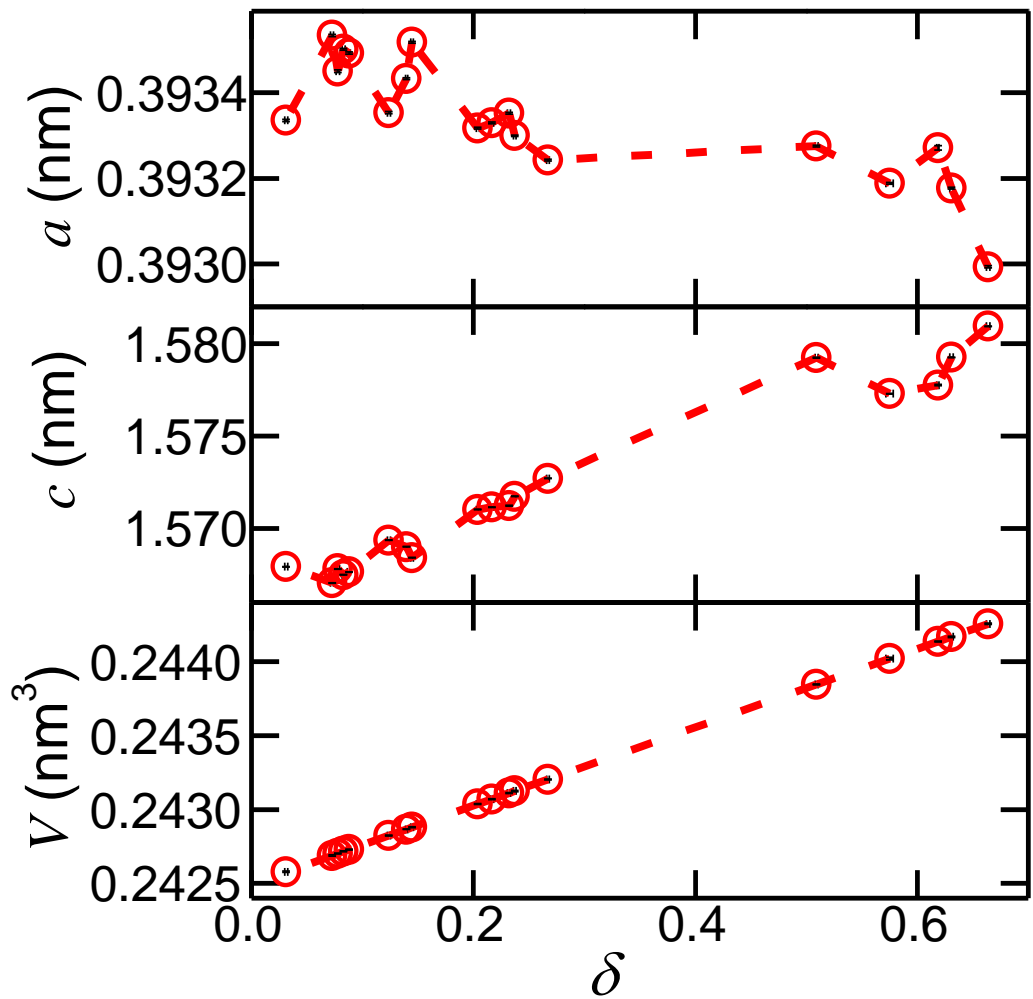

Fig. 3

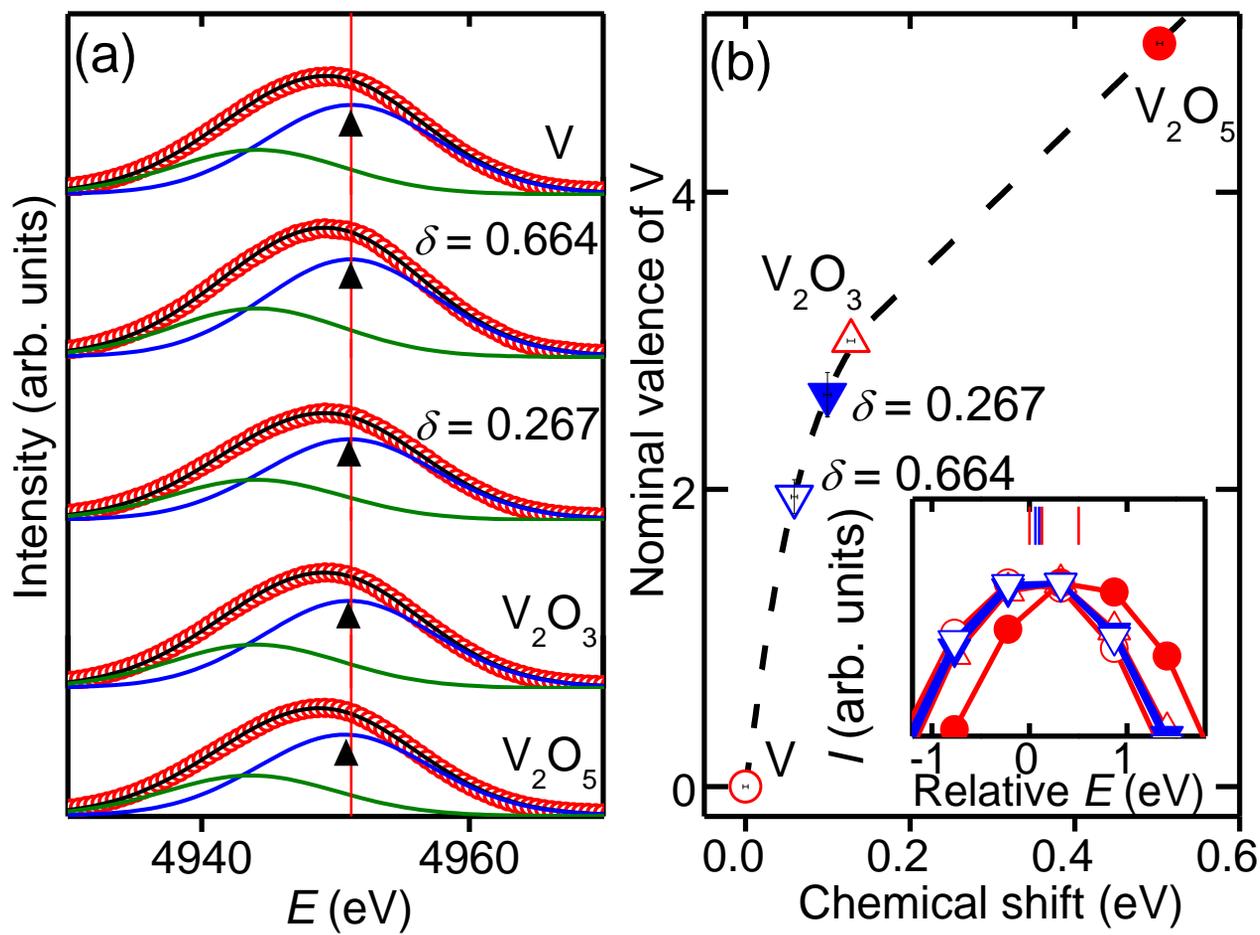

Fig. 4

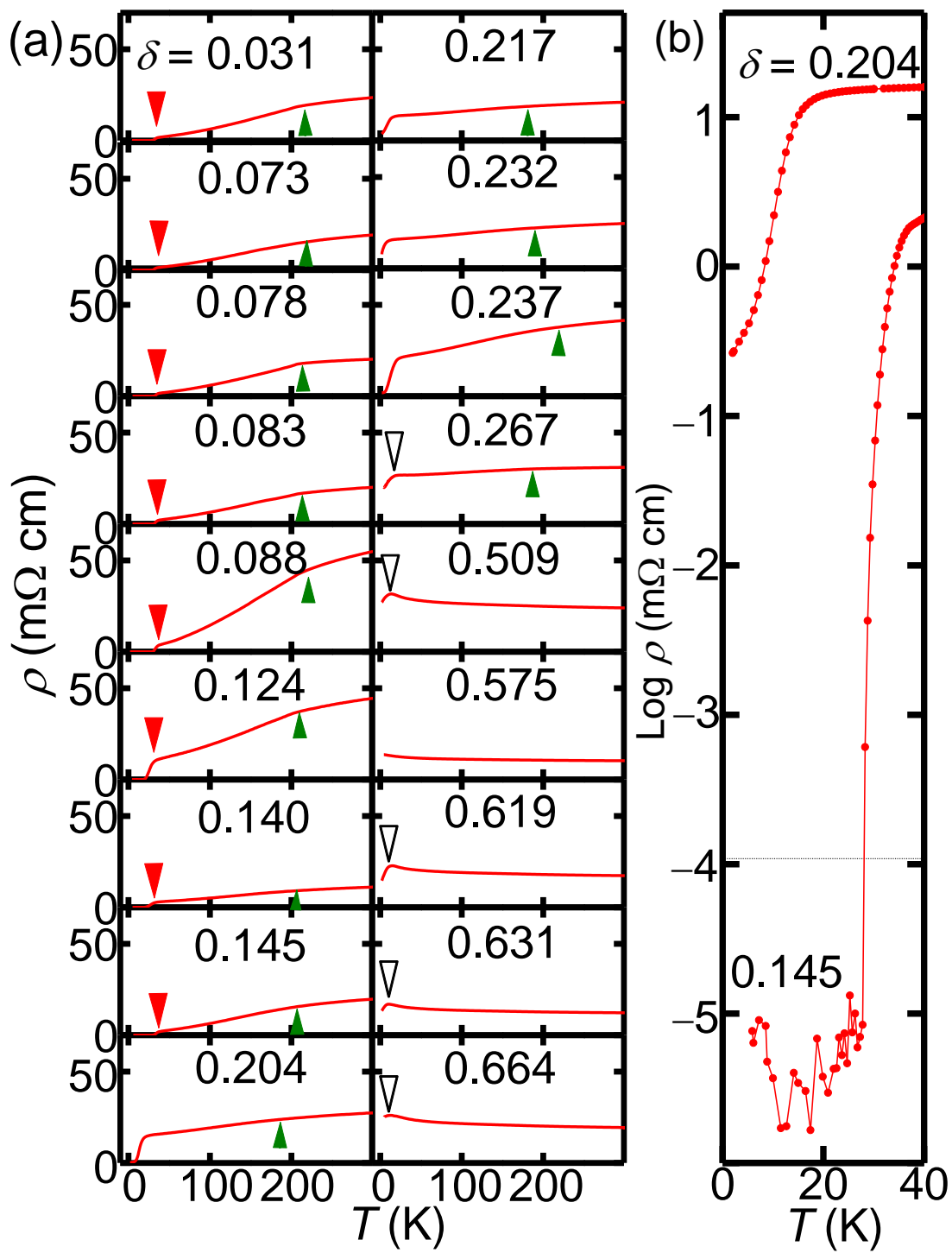

Fig. 5

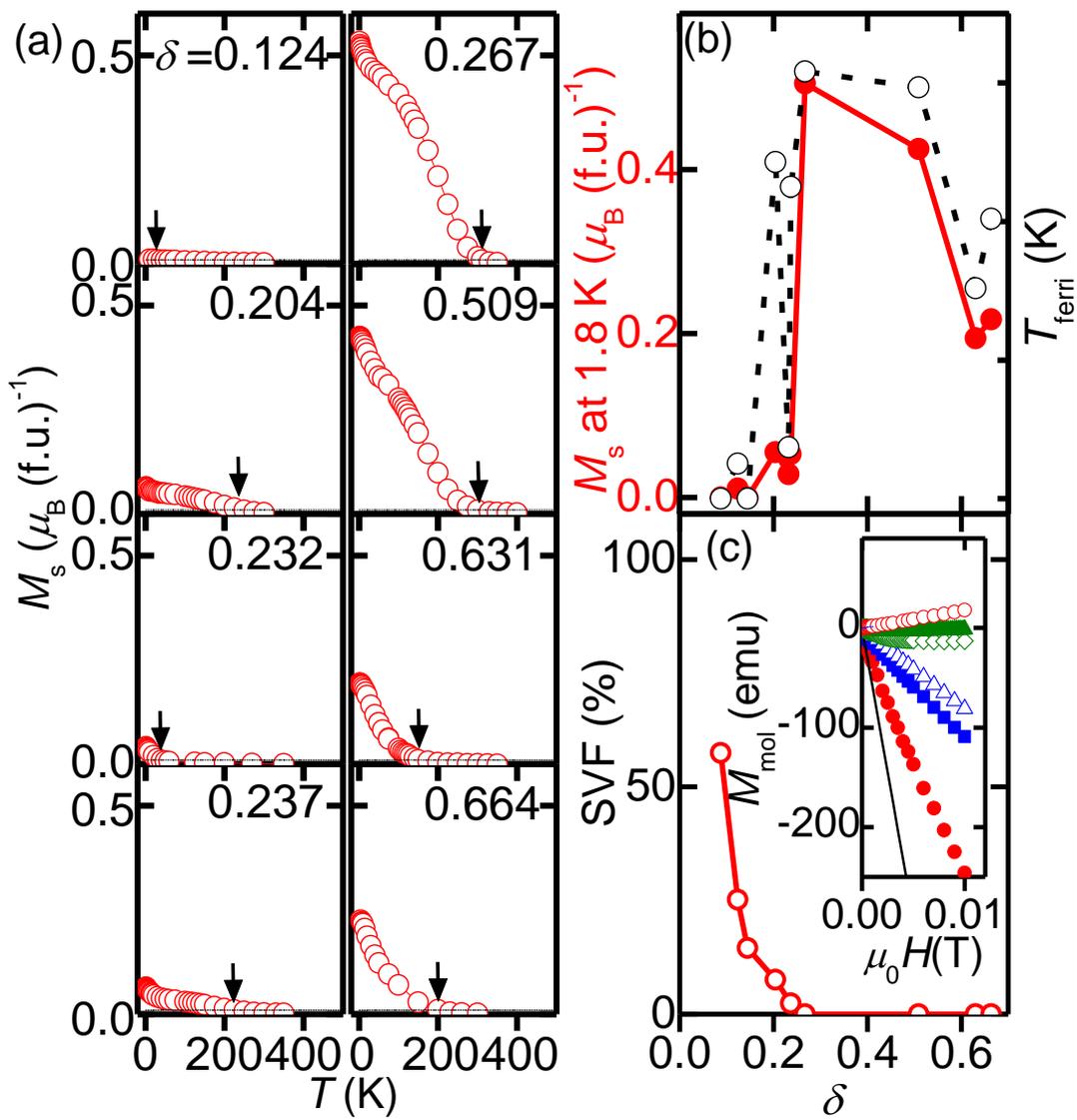

Fig. 6

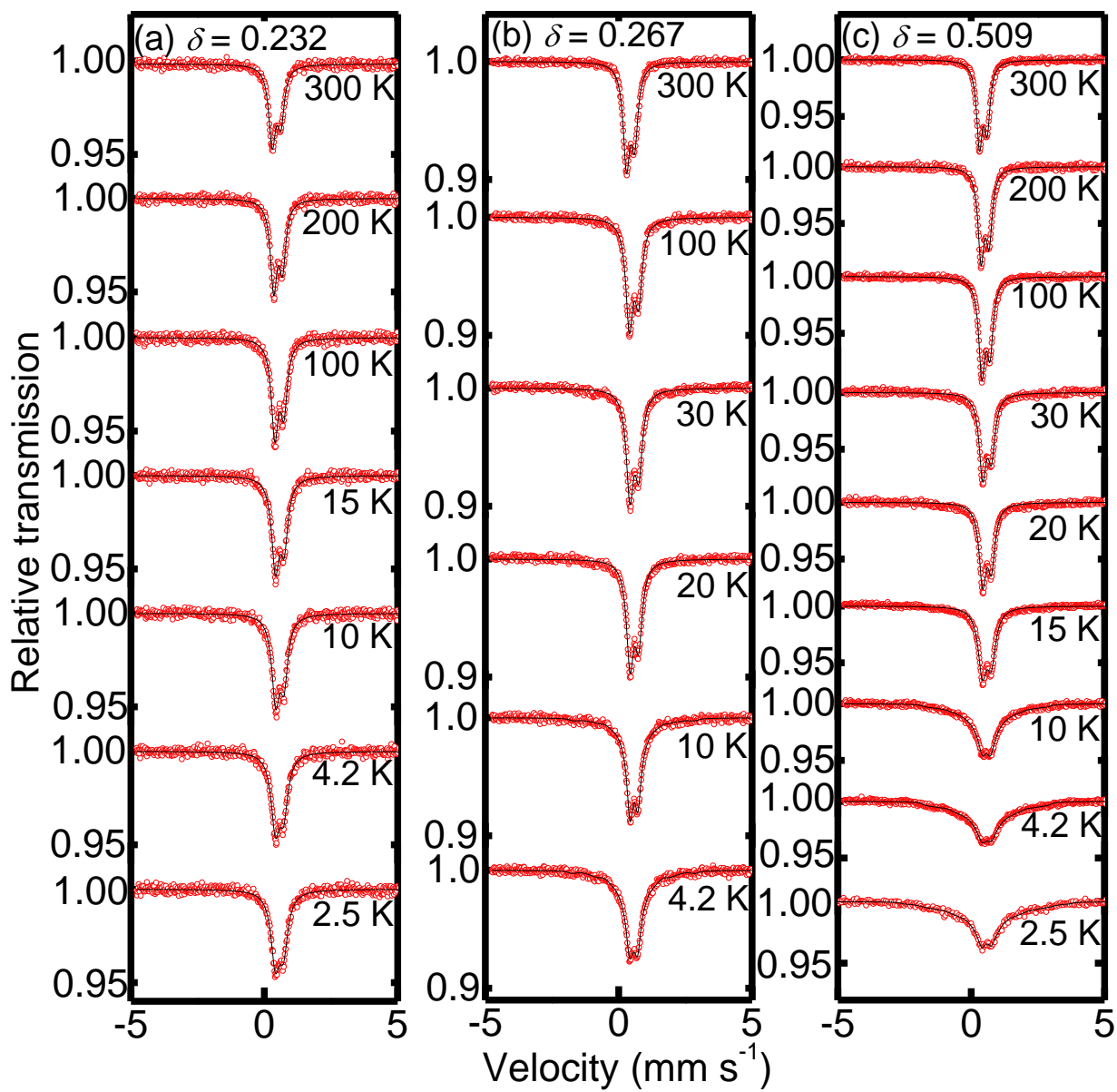

Fig. 7

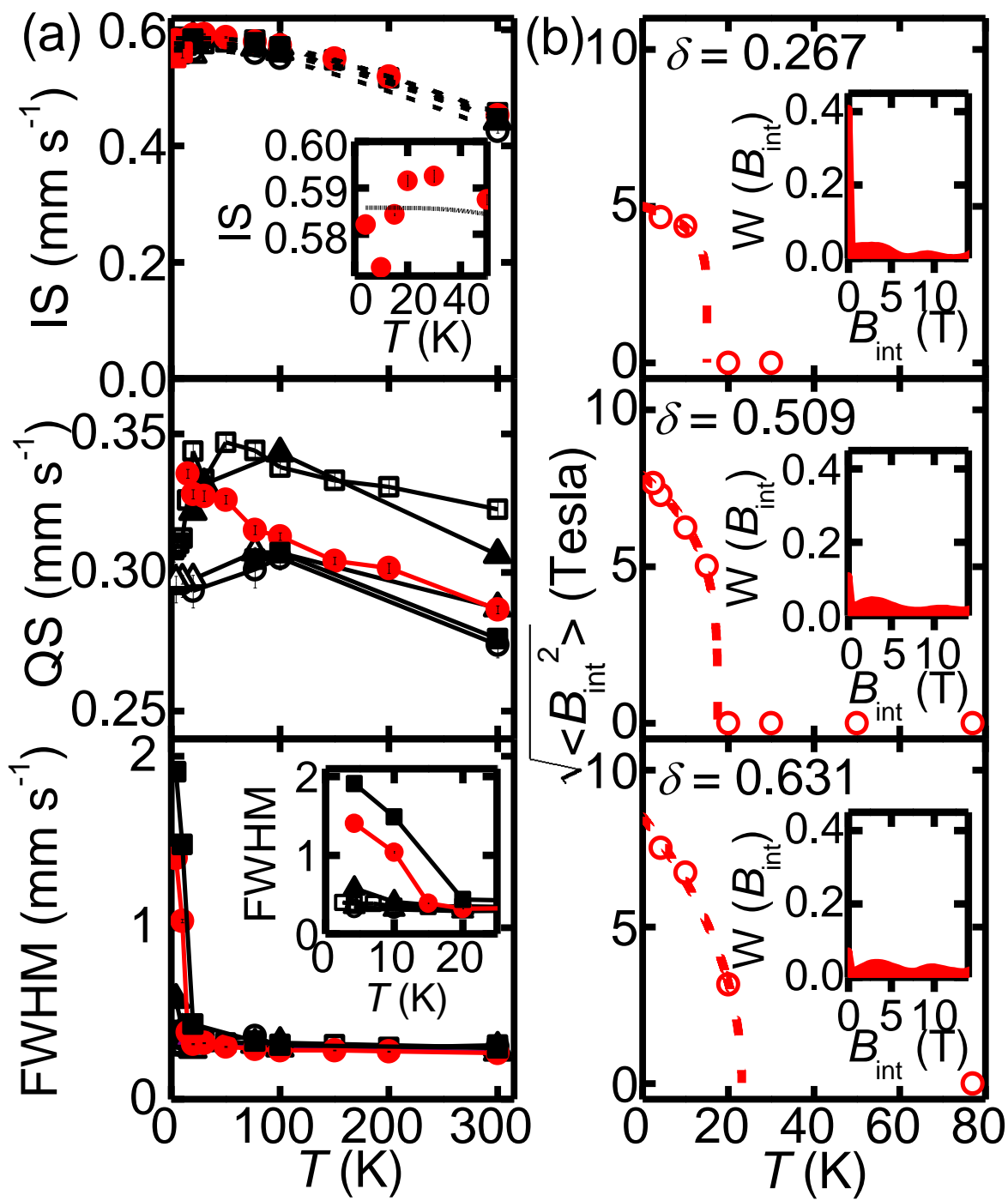

Fig. 8

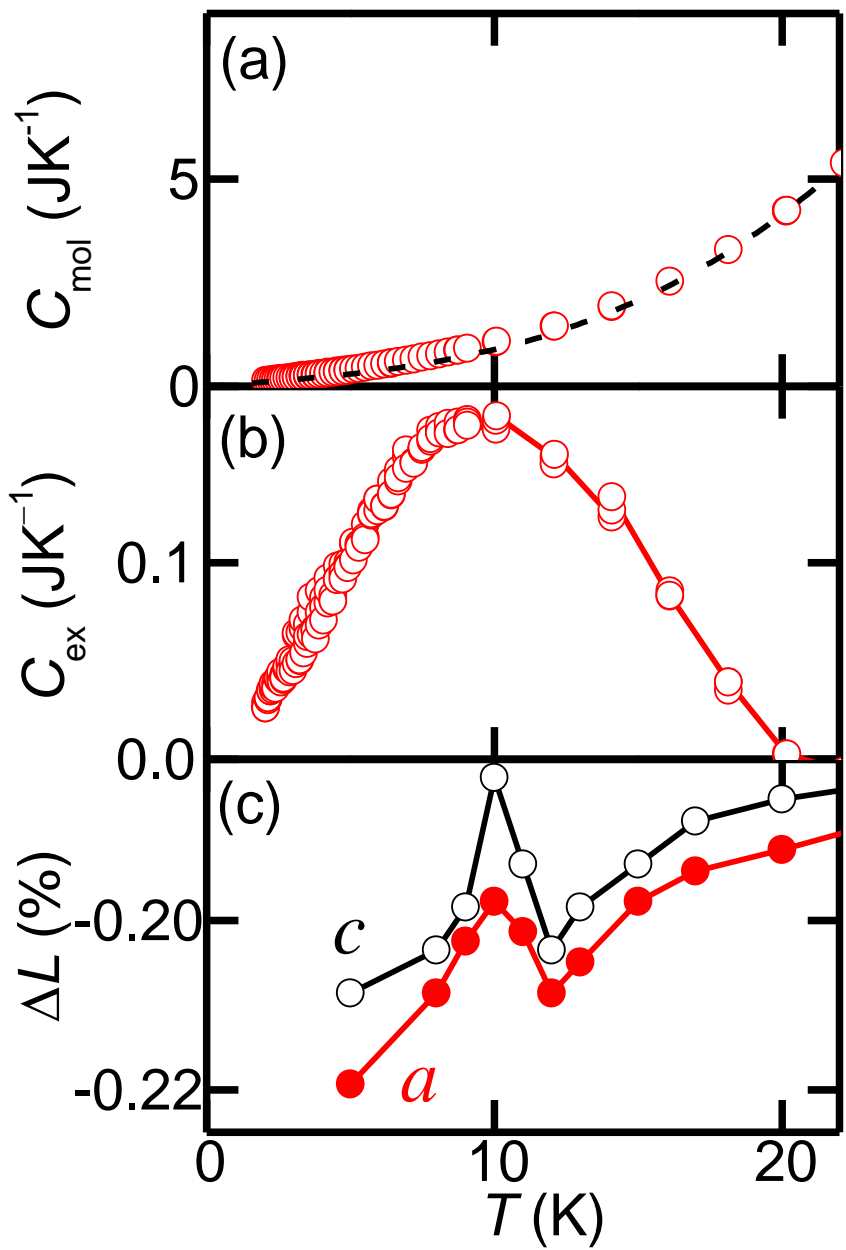

Fig. 9

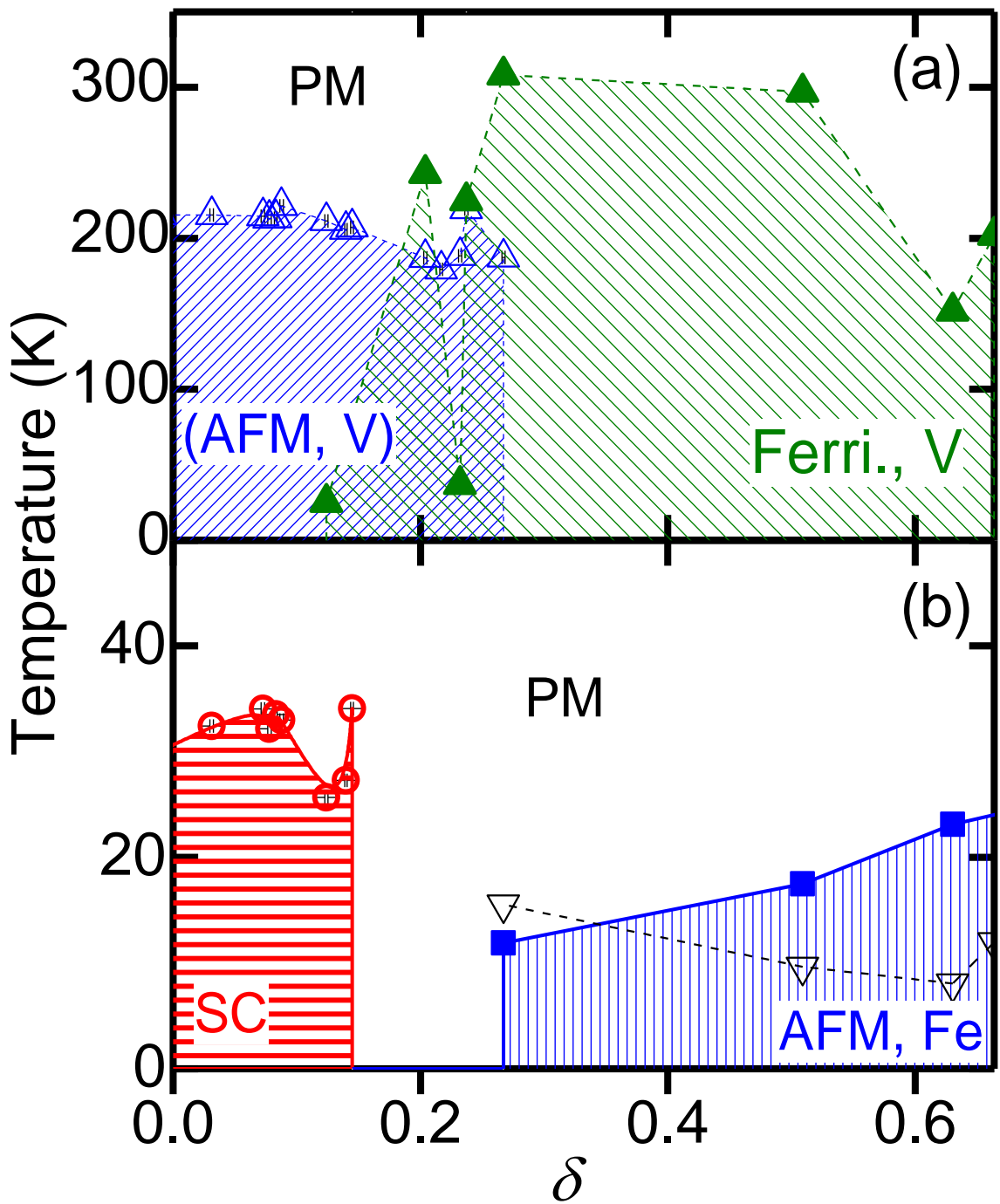

Fig. 10

Tab. 1

| Order | | Moments($\mu_B$) | | $\Delta E$ (meV (f.u.)$^{-1}$) | |
|---|---|---|---|---|---|
| V | Fe | V | Fe | This work | H. Nakamura *et al.* |
| A-AF | PM | 1.9 | 0.0 | 175 - 271 | 766 |
| c-AF | PM | 1.9 | 0.0 | 183 | 58 |
| A-AF | s-AF | 1.9 | 2.2 | 4 - 43 | - |
| c-AF | s-AF | 1.9 | 2.2 | 0 - 44 | 0 |

Tab. 2

| $\delta$ | Order | Moments ($\mu_B$) | | | | $\Delta E$ (meV (f.u.)$^{-1}$) |
|---|---|---|---|---|---|---|
| | | V1 | V2 | V3 | V4 | |
| 0 | A-AF | -1.9 | -1.9 | 1.9 | 1.9 | 0 |
| | FM | 1.9 | 1.9 | 1.9 | 1.9 | 4 |
| | c-AF | -1.9 | 1.9 | -1.9 | 1.9 | 4 |
| | c-AF | 1.9 | -1.9 | -1.9 | 1.9 | 5 |
| | Ferri. | -1.9 | -1.9 | 1.9 | -1.9 | 16 |
| 0.25 | Ferri. | 1.9 | 1.9 | 1.9 | -2.2 | 0 |
| | A-AF (Ferri.) | 1.9 | 1.9 | -1.9 | -2.3 | 7 |
| | FM | 1.9 | 1.9 | 1.9 | 2.3 | 10 |
| 0.50 | c-AF (Ferri.) | 1.9 | -2.3 | 1.9 | -2.3 | 0 |
| | FM | 1.9 | 2.3 | 1.9 | 2.3 | 9 |
| | A-AF | -1.9 | -2.3 | 1.9 | 2.3 | 29 |

**Supplementary information for "Superconducting transition temperatures in Electronic and Magnetic Phase Diagram of a superconductor, $Sr_2VFeAsO_{3-\delta}$"**

Yujiro Tojo, Taizo Shibuya, Tetsuro Nakamura, Koichiro Shoji, Hirotaka Fujioka, Masanori Matoba, Shintaro Yasui, Mitsuru Itoh, Soshi Iimura, Hidenori Hiramatsu, Hideo Hosono, Shigeto Hirai, Wendy Mao, Shinji Kitao, Makoto Seto, and Yoichi Kamihara

## 1. The characterization of crystal structures and chemical compositions for polycrystalline $Sr_2VFeAsO_3$ phase.

Figure S1 shows lattice constants ($a$, $c$) and lattice volumes ($V$) versus nominal oxygen deficiencies ($d$) for polycrystalline $Sr_2VFeAsO_{3-d}$ ($d = -0.1$-0.7) samples. Although there are samples prepared using same nominal compositions, the obtained samples exhibit different $a$, $c$, and $V$. These results indicate that nominal compositions are not equal to intrinsic chemical compositions for $Sr_2VFeAsO_3$ phase. Therefore, an analytical approach is required for determination of the samples' chemical compositions as analyte (54). Provided that the intrinsic oxygen deficiencies correlate linearly to the $V$ of $Sr_2VFeAsO_3$ phase, the dashed line denotes a $d$-$V$ calibration line for intrinsic oxygen deficiency ($\delta$). The $d$-$V$ calibration line is obtained based on three samples showing smaller amount of second phases than other samples. The three samples are $d = 0.15$ with ($a = 0.393353(2)$ nm, $c = 1.56938(1)$ nm, $V = 0.242824(3)$ nm$^3$), $d = 0.20$ with ($a = 0.393299(2)$ nm, $c = 1.57175(2)$ nm, $V = 0.243125(3)$ nm$^3$), and $d = 0.52$ with ($a = 0.393188(2)$ nm, $c = 1.57730(1)$ nm, $V = 0.243845(4)$ nm$^3$).

An inverse function of the $d$-$V$ calibration line is written as $\delta = 377.349$(nm$^{-3}$)$\cdot V - 91.5057$. Figure S2 shows powder XRD patterns of $Sr_2VFeAsO_{3-\delta}$. Each sample contained second phases. Figure S3 shows XRD patterns of $Sr_2VFeAsO_{3-\delta}$ ($\delta = 0.509$) observed and simulated by Rietveld refinement. It is revealed that the total amount of impurity phases is ~7 vol. % for $\delta = 0.509$ sample. Table S1 summarizes crystallographic data of $Sr_2VFeAsO_{3-\delta}$ ($\delta = 0.509$) at room temperature obtained by the Rietveld analysis. (36) The space group is $P4/nmm$. The atomic coordinates are as follows: Sr1 (0.75, 0.75, $z$), Sr2 (0.75, 0.75, $z$), V (0.25, 0.25, $z$), Fe (0.25, 0.75, 0), As (0.25, 0.25, $z$), O1 (0.25, 0.75, $z$), O2 (0.25, 0.25, $z$). The definitions of the two As-Fe-As bond angles $\alpha$, $\beta$ and pnictogen height ($h_{pn}$) in FeAs layer and the V-O1-V bond angle $\gamma$ in $Sr_2VO_{3-\delta}$ layer are illustrated with the tetrahedrons as shown in Fig. 1(b). BVS means the bond valence sum (53), which is obtained from the bond length of between every valence of V and O. Effective valence states of V are less than positive trivalent in $Sr_2VFeAsO_{3-\delta}$.

Valence states of vanadium for $Sr_2VFeAsO_{3-\delta}$ are also quantified based on XRF measurements for V-$K_{\alpha 1}$ level of V, $V_2O_3$, $V_2O_5$, (55-57) and $Sr_2VFeAsO_{3-\delta}$ ($\delta$ = 0.267, 0.664). Oxygen contents of $Sr_2VFeAsO_{3-\delta}$ are quantified based on intensity of O-$K_{\alpha 1}$ emission. Figure S4A shows XRD patterns for V, $V_2O_3$, and $V_2O_5$ as samples used for energy calibration. Figure S4B exhibits XRF spectra for V, $V_2O_3$, $V_2O_5$, and $Sr_2VFeAsO_{3-\delta}$ ($\delta$ = 0.267, 0.664) at Energy ($E$) = 0-15 keV. Although each element of the compounds has been detected by XRF measurements, peak overlap is observed in O-$K_{\alpha 1}$, V-$L_{\alpha 1}$, and V-$L_\beta$ spectra for each sample. As shown in Figure S4C, V-$K_\alpha$ peaks are decomposed into individual spectra (V-$K_{\alpha 1,2}$) using the pseudo-Voigt function (Gaussian-Lorentzian sum function). Valence states of vanadium ions are also determined by peak energies for V-$K_{\alpha 1}$. There is no contradiction between the results of XRF analysis and the results of BVS analysis. A matrix effect (58) gives ten times smaller emission from V and O atoms in $Sr_2VFeAsO_{3-\delta}$ than those from a vanadium metal and vanadium oxides. As shown in Fig. S4D, intensities of XRF spectra in metal V, $V_2O_3$, and $V_2O_5$ could not be simply applied for quantitative analysis on chemical composition of oxygen contents with respect to V in $Sr_2VFeAsO_{3-\delta}$, because the state of bonds of V and O and crystal structure of $Sr_2VFeAsO_{3-\delta}$ are different from those of $V_2O_3$ and $V_2O_5$. The relationship of peak intensities between O-$K\alpha_1$ and V-$L\alpha_1$, V-$L\beta_1$ of the $Sr_2VFeAsO_{3-\delta}$ is in opposite to those of $V_2O_3$ and $V_2O_5$ due to the matrix effect. A semi-quantitative analysis is demonstrated for the oxygen contents. V-$L_{\alpha 1}$, V-$L_{\beta 1}$, and O-$K_{\alpha 1}$ peaks are also decomposed into individual spectra in V, $V_2O_3$, $V_2O_5$, and $Sr_2VFeAsO_{3-\delta}$. Figure S4E shows expanded view of the analysis for $Sr_2VFeAsO_{3-\delta}$. Although a curve is obtained for quantitative analysis on oxygen contents from XRF of V, $V_2O_3$, and $V_2O_5$, the curve could not be used for the quantitative analysis on oxygen contents in $Sr_2VFeAsO_{3-\delta}$. Figure S4F(a) shows the curve.

In this research, we should focus on oxygen contents in same matrix. As shown in Fig. S4F(b), O-$K_{\alpha 1}$ peak's areas, which is defined as relative value to those of V-$K_{\alpha 1}$, show smaller value in $\delta$ = 0.664 than those in $\delta$ = 0.267. The XRF results are in consistent to our characterization using lattice volumes of $Sr_2VFeAsO_{3-\delta}$.

In addition, hydrogen contents were checked by the thermal gas desorption spectrometry (ESCO TDS-1400TV) for $\delta$ = 0.140, 0.267, and 0.664 samples. The hydrogen contents are far smaller than those of deficient oxygen contents in $\delta$ = 0.140, 0.267, 0.664 samples. Although an origin of the hydrogen is controversial for the samples containing second phases, these polycrystalline samples exhibit explicit defects for the oxygen sites in crystallographic phase of $Sr_2VFeAsO_{3-\delta}$.

## 2. Analysis on superconducting properties of $Sr_2VFeAsO_{3-\delta}$

Superconducting properties of polycrystalline $Sr_2VFeAsO_{3-\delta}$ ($\delta = 0.088$) under magnetic field ($H$) are shown in Fig. S5A. Both of $T_c^{onset}$ and $T_c^{offset}$ shift to lower temperature under magnetic field. Figure S5B shows upper critical magnetic field ($H_{c2}$) and interpolated using the empirical parabolic formula. (59) The closed circles are defined by $T_c^{onset}$. The open circles are defined by $T_c^{mid}$. Figure S5C shows magnetization ($M$) as a function of the external magnetic flux ($\mu_0 H$). Clear magnetization hysteresis ($\Delta M$) curves are observed for $\delta = 0.088$ and $0.124$ samples. The superconducting volume fraction defined by magnetic shielding fraction decreases with increasing $\delta$ in $Sr_2VFeAsO_{3-\delta}$. (See Fig. 6c) As shown in Fig. S5D, extended Bean model (60) gives magnetic $J_c$ from the $\Delta M$ as a function of $H$ for the polycrystalline samples. Magnetic $J_c$ is 7.8 kAcm$^{-2}$ at 1.8 K for $\delta = 0.088$ under self-magnetic-field. Magnetic $J_c$s of polycrystalline samples are 100 times smaller than that of single crystalline samples. (61)

## 3. Analysis on $^{57}$Fe Mössbauer spectra for $Sr_2VFeAsO_3$ phase

Figure S6 shows Mössbauer spectra for $Sr_2VFeAsO_{3-\delta}$ ($\delta = 0.124, 0.232, 0.237, 0.267, 0.509,$ and $0.631$).

The analysis of the isomer shift (IS), quadrupole splitting (QS), full width half maximum (FWHM) in the Fig. 8 were performed by the following:

In general, IS($T$) is given by

$$IS(T) = IS_I + IS_{SOD}(T)$$

where $IS_I$ is the intrinsic isomer shift and $IS_{SOD}(T)$ is the second-order Doppler shift which depends on lattice vibrations of the Fe atoms. In terms of the Debye approximation of the lattice vibrations, $IS_{SOD}(T)$ (62) is expressed by the Debye temperature ($\Theta_D$) as

$$\text{IS}_{\text{SOD}}(T) = -E_\gamma \left\{ \frac{3}{2} \frac{k_B T}{Mc^2} \left[ \frac{3}{8} \frac{\Theta_D}{T} + 3 \left(\frac{T}{\Theta_D}\right)^3 \int_0^{\frac{\Theta_D}{T}} \frac{x^3}{e^x - 1} dx \right] \right\}.$$

By fitting the experimental data to the IS($T$) and $IS_{SOD}(T)$, the quantities of $IS_I$ and $\Theta_D$ are obtained as 0.70(1) mm s$^{-1}$ and 454(29) K for the sample $\delta = 0.232$, and 0.68(0) mm s$^{-1}$ and 362(12) K for the sample $\delta = 0.509$, respectively. The decreasing of $IS_I$ with increasing $\delta$ is consistent with following two facts: i. Small $IS_I$ indicates high s-electron densities at the nucleus; ii. High s-electron densities at the nucleus are induced by electron doping due to increasing of $\delta$.

Splitting width of quadrupole interaction energy (QS) is expressed by the energy difference ($\Delta E_q(T)$) between the ground state and the exited state ($\varepsilon$) in QS, the nuclear quadrupole moment ($Q$) and the asymmetry coefficient ($\eta$) as

$$\Delta E_q = 2\varepsilon = \frac{1}{2}eQV_{zz}(1+\frac{\eta^2}{3})^{\frac{1}{2}} = \frac{1}{2}e^2qQ(1+\frac{\eta^2}{3})^{\frac{1}{2}}$$

where the "$2\varepsilon$" indicates the energy difference of the spin angular moment ($I$) between $I_z = \pm\frac{3}{2}$ and $\pm\frac{1}{2}$ in $I = \frac{3}{2}$. The "$\eta$" is equal to $\frac{V_{xx}-V_{yy}}{V_{zz}}$, where $V_{xx}$, $V_{yy}$, $V_{zz}$ denote electric field gradient in $x$, $y$, and $z$ axis. The "$V_{zz}$" is the max electric field gradient and is equals to "$eq$" (63).

The line width of the absorption lines for the samples $\delta$ = 0.267, 0.509, and 0.631 significantly increase at $T$ < 10 K. It indicates that spontaneous magnetic moments of Fe appear in the Fe-sublattice. The Mössbauer spectra and our analysis reveal following experimental results.

i. Debye temperatures of $Sr_2VFeAsO_{3-\delta}$ are about 350-480 K. Oxygen deficiencies decrease the Debye temperature; i.e. the oxygen deficiencies soften $Sr_2VFeAsO_3$ crystallographic phase.

ii. Electric field gradient around the Fe nucleus decreases with increasing oxygen deficiencies. The anomalous kink in the QS-$T$ curve might be due to a transition of the local structure in $Sr_2VFeAsO_3$ phase.

iii. $\delta$ = 0.267, 0.509, and 0.631 samples show magnetic ordered phase of the Fe-sublattice at $T$ < 20 K. The magnetic phase of the Fe-sublattice is an antiferromagnetic (AFM) that might be an magnetic ordered phase like spin density wave involving complex distribution of internal magnetic fields.

## 4. Analysis on temperature dependence of specific heat for polycrystalline $Sr_2VFeAsO_{3-\delta}$ ($\delta$ = 0.509)

Figure S7 shows temperature dependence of molar specific heat ($C_{mol}$) for oxygen deficient $Sr_2VFeAsO_{3-\delta}$ ($\delta$ = 0.509). The dashed line indicates the lattice contribution in the $T$ >> the Debye temperature ($\Theta_D$); i.e. $3R$, $R$: the molar gas constant (65). The inset shows $C_{mol}T^{-1}$ v.s. $T^2$ plots of $Sr_2VFeAsO_{3-\delta}$ ($\delta$ = 0.509). The dashed line indicates the fitted line by the formula $C_{low} = \gamma T + \beta T^3$, where $\gamma$ is Sommerfeld coefficient and $\beta$ is the lattice heat-capacity coefficient.

As shown in Fig. 9, the temperature dependence of the $C_{ex}$ is obtained as the difference between observed $C_{mol}$ and $C_{low}$. $C_{low}$ is an approximate line obtained from Debye model at $T$ << $\Theta_D$. The appearance of $C_{ex}$ is originated from magnetic phase transitions.

## 5. Analysis on relations between $T_c$ and crystallographic parameters for $Sr_2VFeAsO_{3-\delta}$

Figure S8 shows a relation between lattice parameters ($a$, $c$), oxygen deficiency ($\delta$) and superconducting transition temperatures ($T_c$). The onset superconducting transition temperature ($T_c^{onset}$) is determined from the intersection of the two extrapolated lines; one is drawn through the resistivity curve at temperatures in the normal conducting phase just above $T_c$, and the other is drawn through the steepest part of the resistivity curve at temperatures in the superconducting phase. The temperatures of the midpoint of superconducting transitions ($T_c^{mid}$) are determined at the temperatures where the resistivity is 50% of its value at $T_c^{onset}$. The offset temperatures of superconducting transitions ($T_c^{offset}$) are determined from the extrapolated lines in the side of the superconducting state to 0 K. The temperature width of superconducting transition ($\Delta T_c$) is determined from the difference between $T_c^{onset}$ and $T_c^{offset}$. The temperature where the samples exhibit zero resistivity due to superconductivity ($T_c^{zero}$) is determined at the temperature where the resistivity exhibits lower values than an detection limit that is determined from measurement precision as limit to ~1 $\mu\Omega$cm. Maxima of $\rho$-$T$ curves for $\delta \geq 0.267$ are defined as $T_{max}$ for normal conducting samples in Fig. 5. The maxima are denoted by black downward arrows. Clear $T_{anom}$ are determined from $\rho$-$T$ curves for samples with $\delta \leq 0.267$. Table S2 summarize the $T_c$s, $\Delta T_c$s, $T_{max}$s, and $T_{anom}$s for our results.

**Figure captions**

Fig. S1. Lattice constants ($a$, $c$) and lattice volumes ($V$) versus nominal oxygen deficiency (d) for $Sr_2VFeAsO_{3-d}$. Black lines in red plots show standard deviation of the values. The dashed line shows the $d$-$V$ calibration line, based on the nearly stoichiometric samples at $d$ = 0.15, 0.20, 0.52.

Fig. S2. Powder XRD patterns of $Sr_2VFeAsO_{3-\delta}$ (0.031 ≤ $\delta$ ≤ 0.145). The vertical bars at the bottom represent calculated positions of Bragg diffractions of $Sr_2VFeAsO_3$. The black, blue, green, and orange arrows represent Bragg diffractions due to impurity phases $SrV_2O_6$, $Sr_2VO_4$, FeAs and $Fe_3O_4$.

Fig. S3. XRD patterns of $Sr_2VFeAsO_{3-\delta}$ ($\delta$ = 0.509) observed (red circles) and simulated by Rietveld analysis (green line). The gray line represents the difference between the two. The vertical bars at the bottom represent the calculated positions of Bragg diffractions of $Sr_2VFeAsO_{2.5}$ (black), $SrV_2O_6$ (blue) and $Fe_3O_4$ (orange). A ratio of the $SrV_2O_6$ phase and $Fe_3O_4$ phase are 6.9 vol.% and 0.3 vol.%.

Fig. S4. (Color online) The details of XRF measurements for valence state of V in $Sr_2VFeAsO_{3-\delta}$.
(A) Powder XRD patterns of vanadium and vanadium oxides. The vertical bars at the bottom represent calculated positions of Bragg diffractions of V, $V_2O_3$, $V_2O_5$, respectively. (B) Observed XRF spectrum of $Sr_2VFeAsO_{3-\delta}$ ($\delta = 0.267, 0.664$), vanadium and vanadium oxides. The intensity of V-K$\alpha$ spectra (55-57) of $Sr_2VFeAsO_{3-\delta}$ ($\delta = 0.267, 0.664$) are ~10 times smaller than those of $V_{metal}$, $V_2O_3$ and $V_2O_5$, indicating matrix effect (58) due to X-ray absorption of other chemical species. (C) Observed (red circles) and fitted (black solid line) V-K$\alpha_1$ (blue solid line) and V-K$\alpha_2$ (green solid line) spectra for $Sr_2VFeAsO_{3-\delta}$ ($\delta = 0.267, 0.664$), vanadium metal ($V_{metal}$)) and vanadium oxides ($V_2O_3$, $V_2O_5$). The orange solid line represents the difference between the observed and fitted spectrum. (D) Observed (red circles) and fitted (black solid line) (O-K$\alpha_1$ (blue solid line), V-L$\alpha_1$ (green solid line) and V-$\beta_1$ (purple solid line)) spectra for $Sr_2VFeAsO_{3-\delta}$ ($\delta = 0.267, 0.664$), vanadium metal ($V_{metal}$)) and vanadium oxides ($V_2O_3$, $V_2O_5$). The black dashed line represents the background, and the orange solid line represents the difference between the observed and fitted spectrum. (E) Expanded view of O-K$\alpha_1$ (blue solid line), V-L$\alpha_1$ (green solid line) and V-$\beta_1$ (purple solid line) XRF spectra for $Sr_2VFeAsO_{3-\delta}$ ($\delta = 0.267, 0.664$) in Fig. S4D. (F) Relationship between relative area of O-K$\alpha_1$ and oxygen contents in V, $V_2O_3$, $V_2O_5$ and $Sr_2VFeAsO_{3-\delta}$ ($\delta = 0.267, 0.664$). (a) Relative peak area ratio of O-K$\alpha_1$ in V, $V_2O_3$ and $V_2O_5$ based on that of $V_2O_5$ as a function of nominal O contents based on a V element. The dashed line indicates a calibration curve of oxygen contents in vanadium oxides. (b) Relative peak area ratio of O-K$\alpha_1$ in $Sr_2VFeAsO_{3-\delta}$ ($\delta = 0.267$, 0.664), based on that of the sample $\delta = 0.267$. Indeed, oxygen contents decrease with increasing $\delta$.

It should be noted that there are difficulties to determine quantitative oxygen contents of $Sr_2VFeAsO_{3-\delta}$ with XRF. Primary reasons are following: (i) The energy of V-L$\alpha_1$, V-L$\beta_1$, and O-K$\alpha_1$ are very close in the range of 0.01 keV, which is detection limit of WDX, and the observed V-L$\alpha_1$, V-L$\beta_1$ and O-K$\alpha_1$ spectrum largely overlapped. Therefore, provided the spectrum of them are divided by least squares fitting, the arbitrariness of fitting inevitably occurs. (ii) The calibration of oxygen contents with $V_{metal}$, $V_2O_3$ and $V_2O_5$ (Fig. S4F(a)) could not been applied to determine chemical composition of O with respect to V in $Sr_2VFeAsO_{3-\delta}$, bacause the state of bonds of V and O and crystal structure of $Sr_2VFeAsO_{3-\delta}$ are different from those of $V_2O_3$ and $V_2O_5$. (iii) Because of the matrix effect, the relationship of peak intensity or peak area between O-K$\alpha_1$ and V-L$\alpha_1$, V-L$\beta_1$ of the samples $Sr_2VFeAsO_{3-\delta}$ is opposite to that of $V_2O_3$ and $V_2O_5$ (See Fig. S4D). (iv) Although it was possible to distinguish crystallographic phase of $Sr_2VFeAsO_{3-\delta}$ and that of impurity phase with Electron Probe Micro Analyzer (EPMA), it was difficult to detect O-K$\alpha_1$ spectrum with superior S/N ratio with EPMA, because the measurement range of $Sr_2VFeAsO_{3-\delta}$ crystals is ~ μm order and the sufficient intensity of O-K$\alpha_1$ spectrum could been hardly detected.

Fig. S5. The superconducting properties of $Sr_2VFeAsO_{3-\delta}$ under magnetic fields. (A) Temperature ($T$) dependences of electrical resistivity ($\rho$) of $Sr_2VFeAsO_{3-\delta}$ ($\delta = 0.088$) in the low temperature region from 2 K to 60 K under different magnetic fields from $\mu_0H = 0$-9 T. The inset shows the $\rho$-$T$ curve below $T = 40$ K. In the inset, the red dashed denote the detection limit of our measurement. With increasing magnetic fields, the onset superconducting transition temperature slowly shifts towards lower temperature, however the transition width significantly gets broad. The $\rho$ of the sample decreases below our detection limit under 0 T, on the other hand, the $\rho$ exhibits finite values at 2.5 K above the detection limit under above 0.2 mT. This disappearance of zero resistivity under magnetic fields is possibly due to the existence of impurity phases in the grain boundary in the sample. (B) The temperature ($T$) dependences of the upper critical fields ($\mu_0H_{c2}$) at $T_c^{onset}$ (closed red circles) and $T_c^{mid}$ (open blue circles) of $Sr_2VFeAsO_{3-\delta}$ [$\delta = 0.088$]. The $\mu_0H_{c2}$ ($T$) near $T_c^{onset}$ was determined from the onset transitions of the $\rho$-$T$ curves under every magnetic field. A rapid increase of $\mu_0H_{c2}$ ($T$) near $T_c^{onset}$ with decreasing temperature is observed and $\frac{dHc2}{dT}|_{T=T_c^{onset}}$ from 1 T to 9 T is ~ -9.3 TK$^{-1}$, which is a smaller slope than that of a nominal $\delta = 0$ sample reported by Zhu et al (7). The $\mu_0H_{c2}$ ($T$) near $T_c^{mid}$ was determined from a criterion of the value of 50%$\rho$ at $T_c^{onset}$ under every magnetic field. $\frac{dHc2}{dT}|_{T=T_c^{mid}}$ from 1 T to 9 T is ~ −1.3 TK$^{-1}$, which is also a smaller slope than that of nominal $\delta = 0.1, 0.3$ and $0.5$ samples reported by Han et al (20). These $\mu_0H_{c2}$ -$T$ curves were obtained by an empirical parabolic formula $H_{c2}$ ($T$) = $H_0 \left\{ 1 - \left(\frac{T}{T_c}\right)^2 \right\}$ (59). Then, $\mu_0H_{c2}$ (0) (=$\mu_0H_0$) near $T_c^{onset}$ and $T_c^{mid}$ of the $\delta = 0.088$ sample are ~240 T and ~20 T. The irreversibility fields ($\mu_0H_{irr}$)-$T$ curve of the sample could not be obtained, because the zero resistivity did not be observed at 2.5 K above

0.2 mT. (C) Magnetic field ($\mu_0H$) dependence of magnetization ($M$) of the samples Sr$_2$VFeAsO$_{3-\delta}$ [$\delta$ = 0.088 and 0.124] at 1.8 K (red closed circles) and at 10 K (blue open circles), respectively. The inset shows expanded the molar magnetism ($M_{mol}$)-$\mu_0H$ curves in the range from 0 T to 10$^{-2}$ T. The black dashed lines denote the $M_{mol}$ of the perfect diamagnetism. The sample $\delta$ = 0.088 exhibits the rather larger magnetic hysteresis $\Delta M$, whose value is ~1.1 $\mu_B$ (formula unit (f.u.))$^{-1}$ under 0 T at 1.8 K. On the other hand, the sample $\delta$ = 0.124 exhibits the rather smaller $\Delta M$, whose value is ~0.064 $\mu_B$ (f.u.)$^{-1}$ under 0 T at 1.8 K. (D) Magnetic field ($\mu_0H$) dependence of magnetic critical current density ($J_c$) of the samples Sr$_2$VFeAsO$_{3-\delta}$ [$\delta$ = 0.088, 0.124] at 1.8 K (closed red circles), 4.2 K (open blue circles), 10 K (closed green triangles) and 20 K (open black triangles), respectively. As shown in Fig. S5C, the magnetic $J_c$ is obtained from the observed magnetization hysteresis loops based on the extended Bean model (60) as $J_c = \frac{20 \Delta M}{t(1-\frac{t}{3l})}$, where $l$ and $t$ are length of the rectangular shape ($l > t$). Units for the variables are summarized as $\Delta M$ (emu cm$^{-3}$ or 10$^3$Am$^{-1}$), magnetic $J_c$ (Acm$^{-2}$), $t$ (cm), and $l$ (cm). The $\delta$ = 0.088 sample exhibits $J_c$ ~7.8 kAcm$^{-2}$ and ~1.6 kAcm$^{-2}$ under 0 T and 5 T at 1.8 K, respectively. Magnetic $J_c$ values are 100 times smaller than those of a single crystalline sample reported by Katagiri et al (61). The $\delta$ = 0.124 sample exhibits $J_c$~3.6 kAcm$^{-2}$ and ~0.11 kAcm$^{-2}$ under 0 T and 5 T at 1.8 K, respectively. The slight increase of oxygen deficiency tremendously weakens the magnetic $J_c$. The magnetic $J_c$ rapidly decreases with increasing temperatures. The values of the magnetic $J_c$ of both samples at 10 K are 10 times smaller than those of them at 1.8 K.

Fig. S6. $^{57}$Fe Mössbauer spectrum of the samples Sr$_2$VFeAsO$_{3-\delta}$ ($\delta$ = 0.124, 0.232, 0.237, 0.267, 0.509 and 0.631) at several temperatures described in the figure. The solid lines are fitted patterns.

Fig. S7. The temperature ($T$) dependence of observed molar heat capacity ($C_{mol}$) of Sr$_2$VFeAsO$_{3-\delta}$ ($\delta$ = 0.509). The dashed line indicates the lattice contribution at $T \gg$ the Debye temperature ($\Theta_D$); i.e. 3R, R: the molar gas constant (65). The inset shows $C_{mol}T^{-1}$ v.s. $T^2$ plots of Sr$_2$VFeAsO$_{3-\delta}$ ($\delta$ = 0.509). The dashed line indicates the fitted line by the formula $C_{low} = \gamma T + \beta T^3$, where $\gamma$ is Sommerfeld coefficient and $\beta$ is the lattice heat-capacity coefficient at $T \ll \Theta_D$.

Fig. S8. The relationship between lattice parameters ($a$, $c$), oxygen deficiency ($\delta$) and superconducting transition temperatures ($T_c$). (A) Oxygen deficiency ($\delta$) dependences of lattice constants $a$ (a) and $c$ (b) and $T_c^{onset}$ (open red circles) and $T_c^{mid}$ (closed red circles) (c). (B) Ratio of lattice constants $a$ and $c$ ($a\,c^{-1}$) dependences of. $T_c$ ($T_c^{onset}$: open circles, $T_c^{mid}$: closed circles, $T_c^{zero}$: closed triangles). In both figures, the area enclosed red line indicates bulk superconducting (SC) phase, black line in red plots show standard deviation of the values and the arrows indicate $\delta$ giving the highest $T_c^{onset}$ and $T_c^{mid}$ of all of the prepared samples.

Table S1. Crystallographic data of $Sr_2VFeAsO_{3-\delta}$ ($\delta = 0.509$) at room temperature by Rietveld analysis. The space group is *P4/nmm*. The atomic coordinates are as follows: Sr1(0.75, 0.75, $z$), Sr2(0.75, 0.75, $z$), V(0.25, 0.25, $z$), Fe(0.25, 0.75, 0), As(0.25, 0.25, $z$), O1(0.25, 0.75, $z$), O2(0.25, 0.25, $z$). The definitions of the two As-Fe-As bond angles $\alpha$, $\beta$ and pnictogen height ($h_{pn}$) in FeAs layer and the V-O1-V bond angle $\gamma$ in $Sr_2VO_{3-\delta}$ layer are illustrated with the tetrahedrons as shown in Fig. 1(b). BVS means the bond valence sum (53), which is obtained from the bond length of between every valence of V and O.

Table S2. Superconducting transition temperatures and anomalous kinks of the prepared samples of $Sr_2VFeAsO_{3-\delta}$. The onset superconducting transition temperature ($T_c^{onset}$) is determined from the intersection of the two extrapolated lines; one is drawn through the resistivity curve in the normal state just above $T_c$, and the other is drawn through the steepest part of the resistivity curve in the superconducting state. The temperature of the midpoint of monotonic decrease of resistivity ($T_c^{mid}$) is determined at the temperature where the resistivity is 50% of its value at $T_c^{onset}$. The offset temperature of superconducting transition ($T_c^{offset}$) is determined from the extrapolated line in the side of the superconducting state to 0 K. The temperature width of superconducting transition ($\Delta T_c$) is determined from the difference between $T_c^{onset}$ and $T_c^{offset}$. The temperature where the samples exhibit zero resistivity due to superconductivity ($T_c^{zero}$) is determined at the temperature where the resistivity exhibits the our detection limit, which is determined from measurement precision as limit to ~1 $\mu\Omega$. The The anomalous maxima ($T_{max}$) and the anomalous kink ($T_{anom}$) in the $\rho$-$T$ curves are determined from the intersection of the two extrapolated lines of the lower and upper curves near the maxima and kink, respectively. The "p" indicates that anomaly kink possibly exists at temperatures < ~10 K.

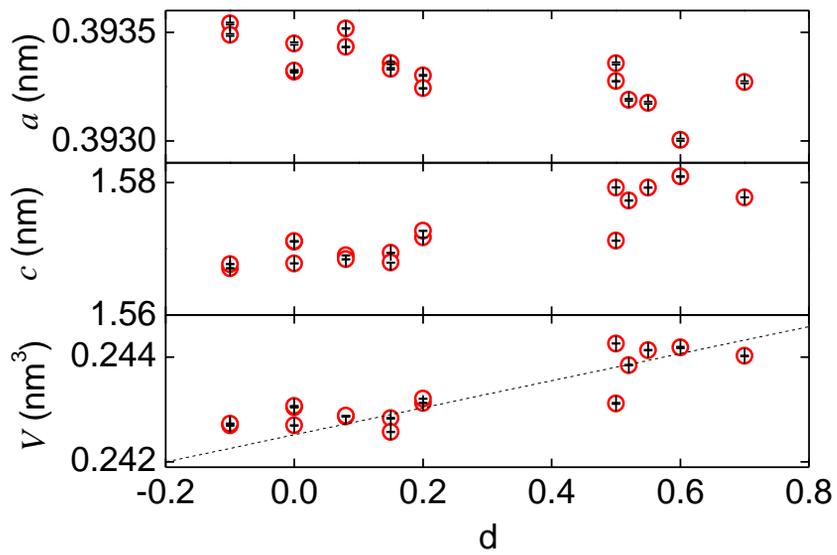

Fig. S1

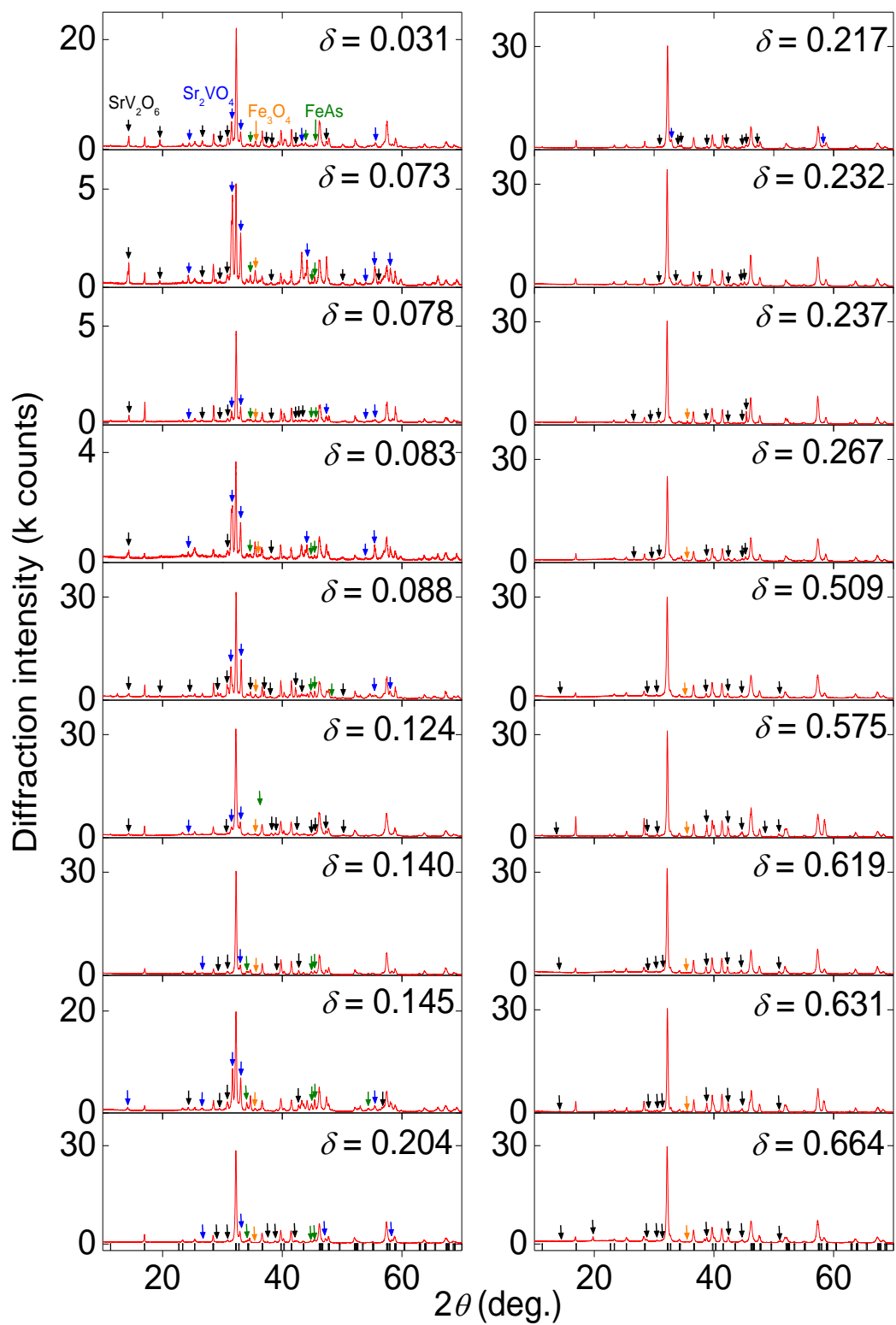

Fig. S2

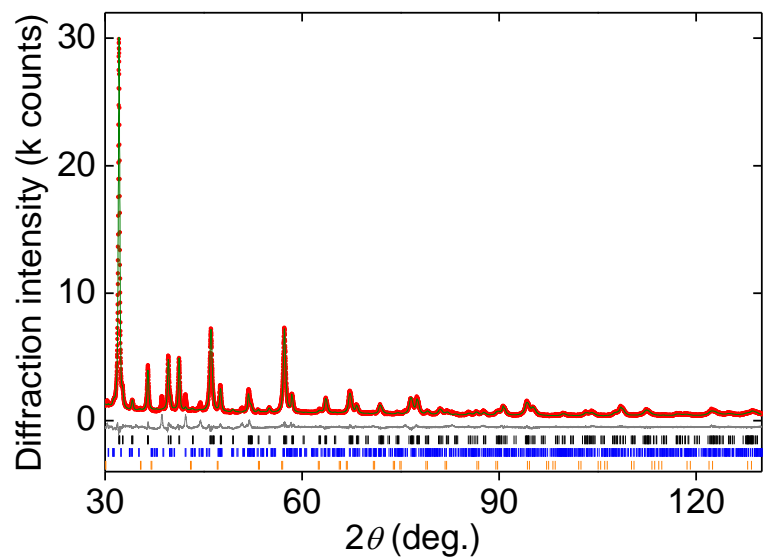

Fig. S3

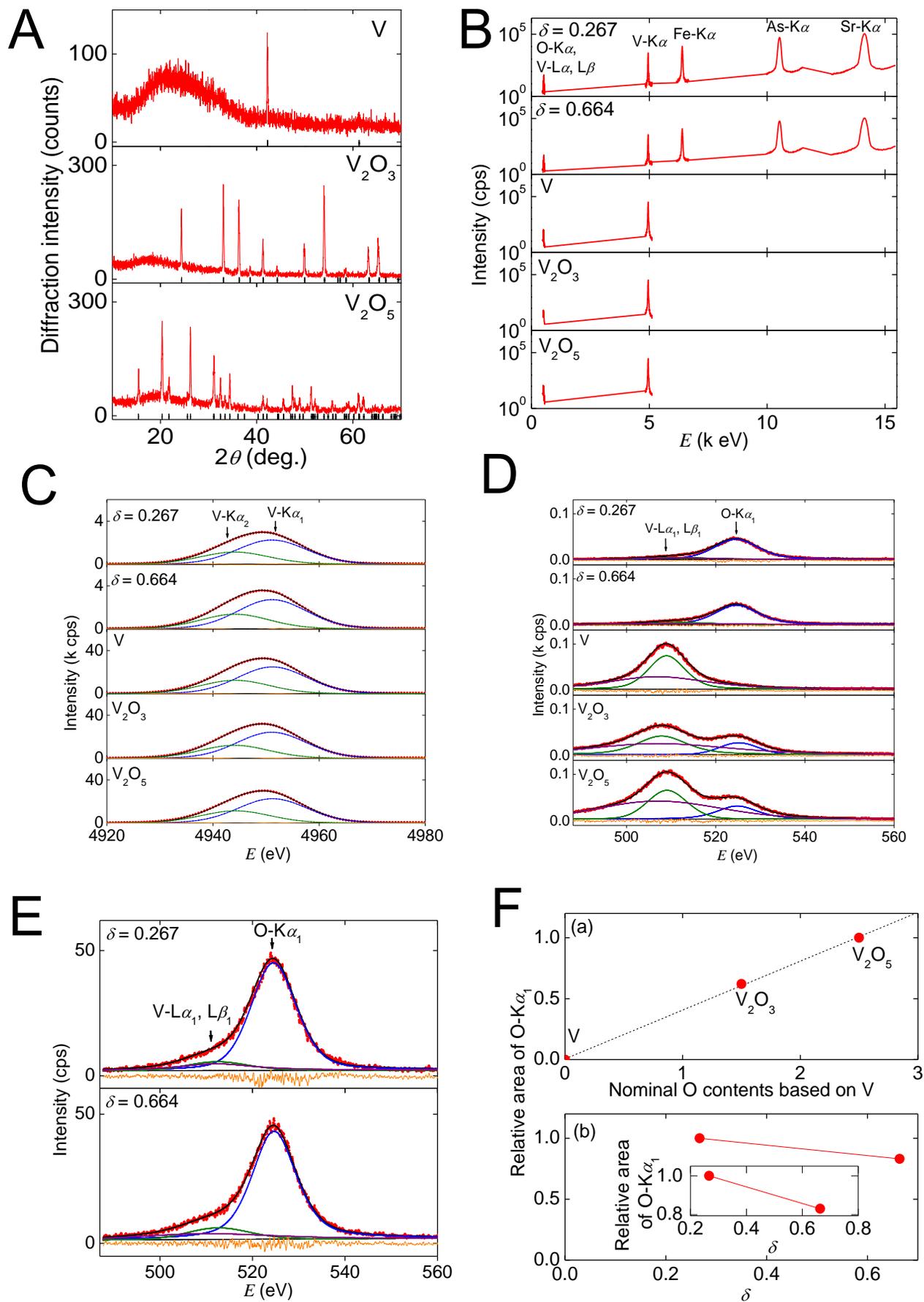

Fig. S4

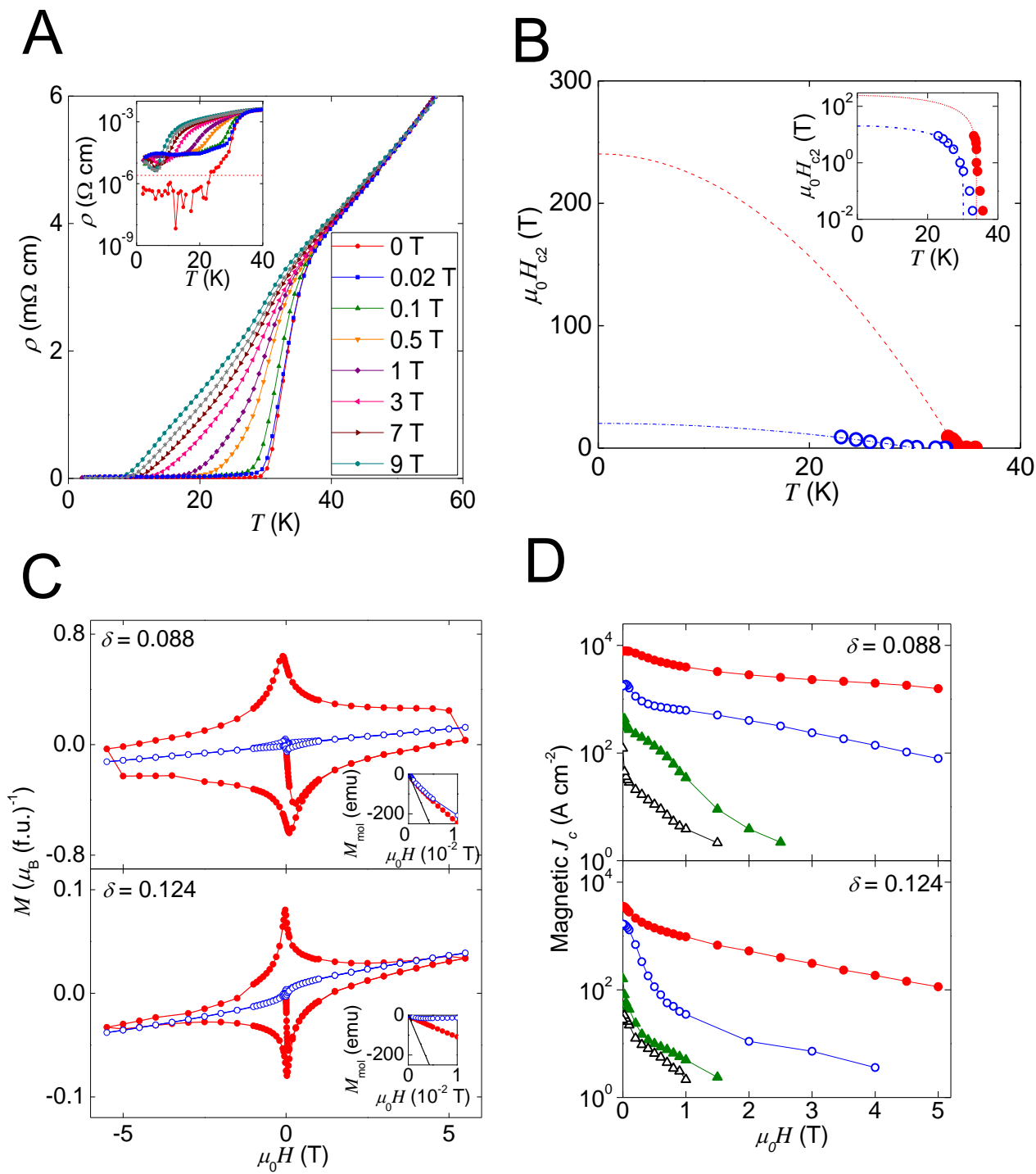

Fig. S5

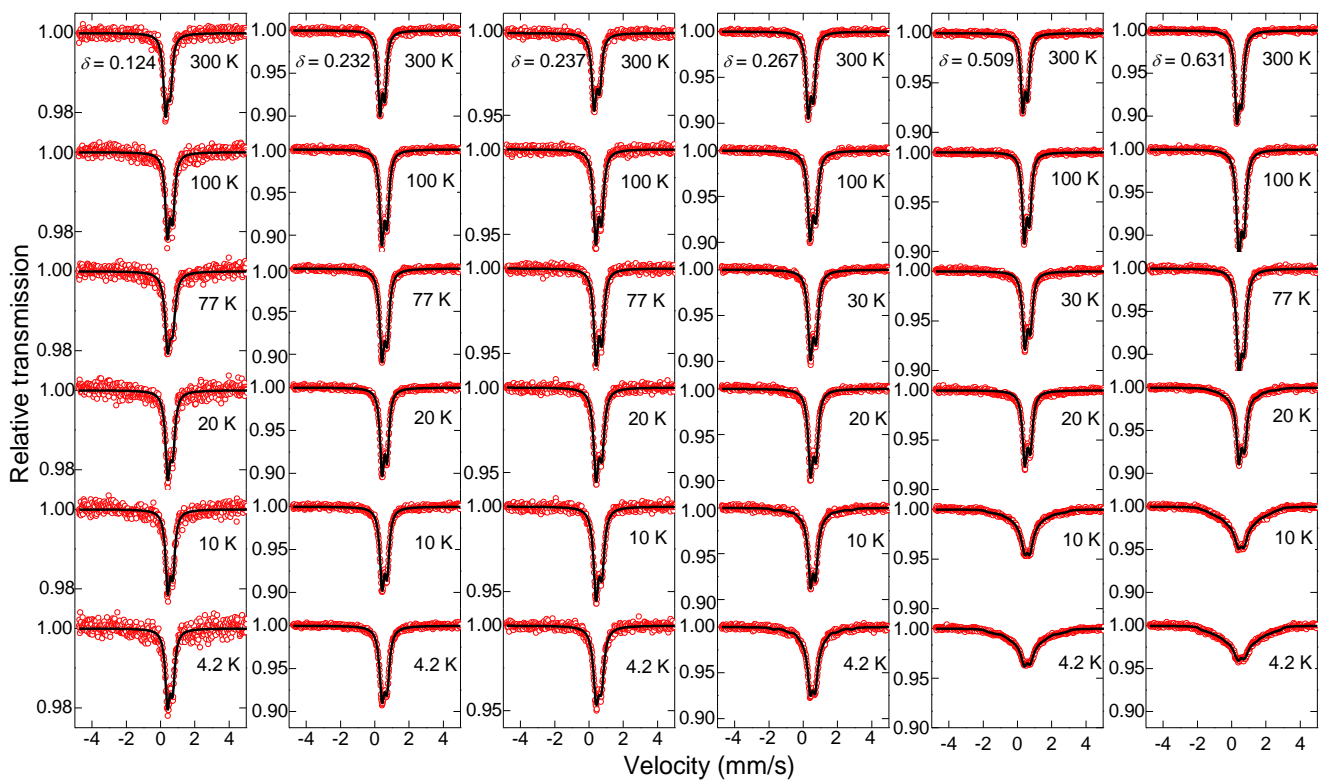

Fig. S6

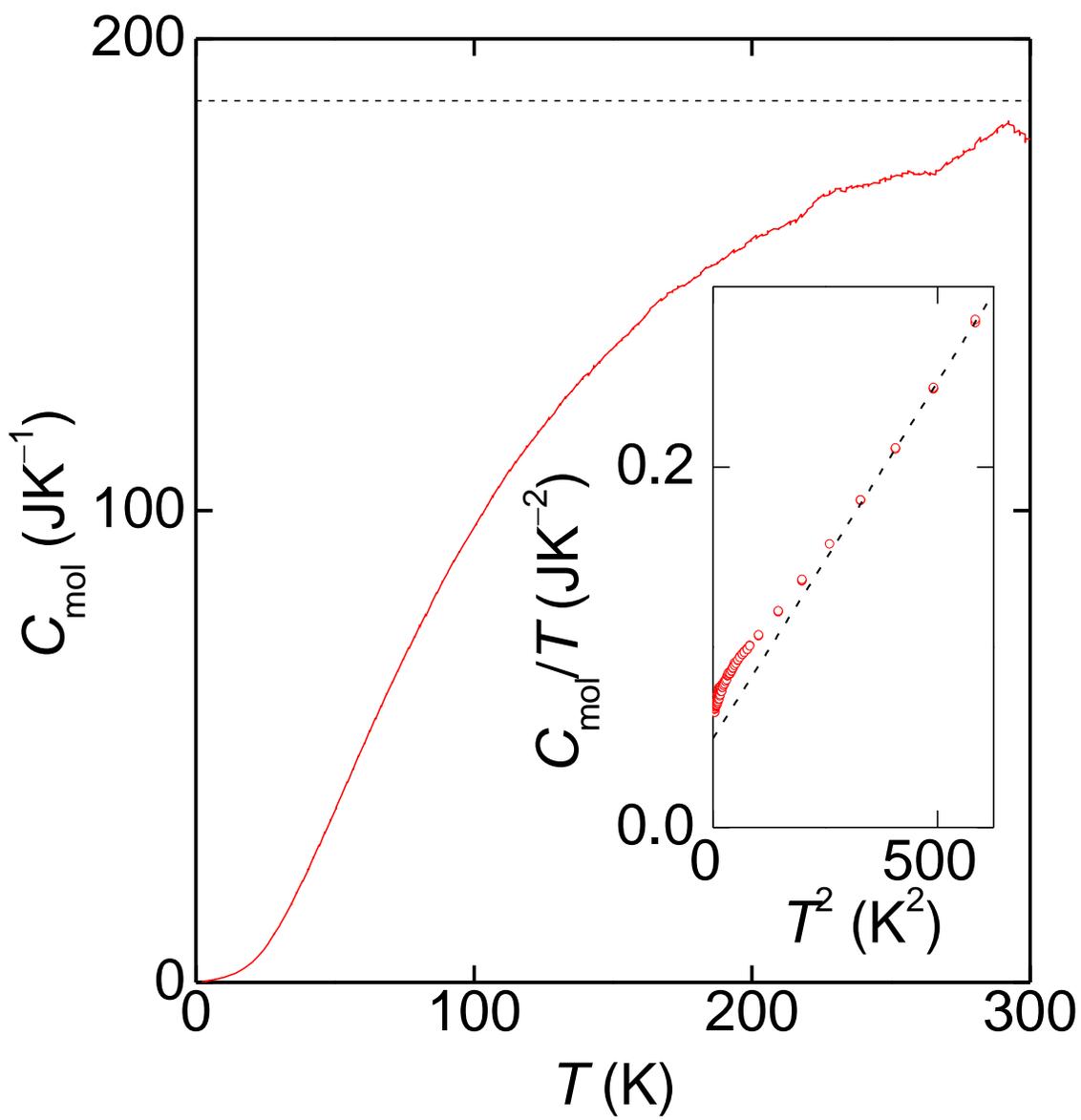

Fig. S7

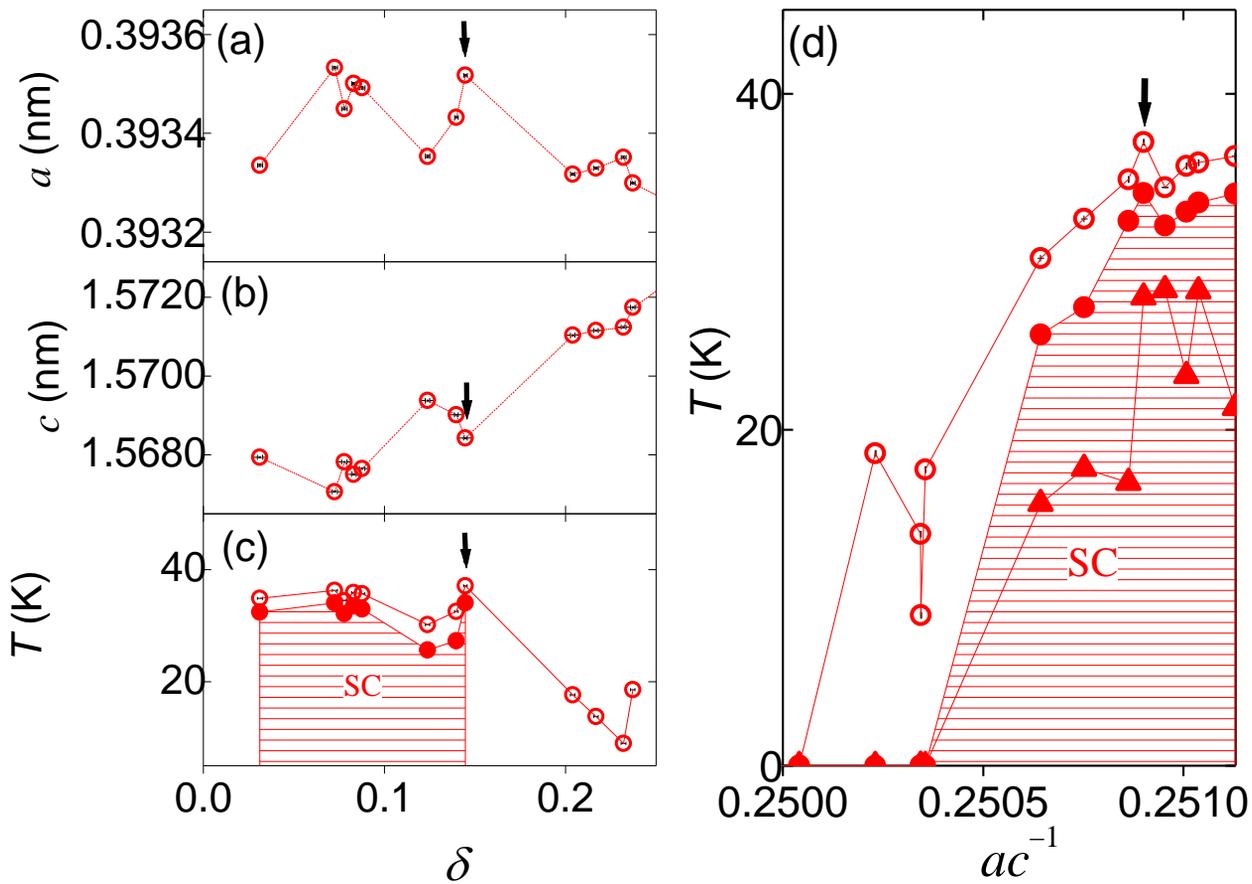

Fig. S8

# Table S1

|   |   |   |
|---|---|---|
| $a$ (nm) | | 0.393188(2) |
| $c$ (nm) | | 1.57730(1) |
| $V$ (nm³) | | 0.243845(4) |
| $z$ | Sr1 | 0.1914(1) |
| | Sr2 | 0.4128(1) |
| | V | 0.31077(0) |
| | Fe | 0 |
| | As | 0.0931(1) |
| | O1 | 0.2966(5) |
| | O2 | 0.4407(7) |
| $BVS_V$ for $V^{3+}$ | | 2.56 |
| $BVS_V$ for $V^{4+}$ | | 2.86 |
| $BVS_V$ for $V^{5+}$ | | 3.01 |
| $\alpha$ (deg.) | | 106.43(9) |
| $\beta$ (deg.) | | 110.01(8) |
| $\gamma$ (deg.) | | 166.7(4) |
| $h_{pn}$ (nm) | | 0.146990(2) |

# Table S2

| $\delta$ | $T_c^{onset}$ (K) | $T_c^{mid}$ (K) | $T_c^{offset}$ (K) | $\Delta T_c$ (K) | $T_c^{zero}$ (K) | $T_{max}$ (K) | $T_{anom}$ (K) |
|---|---|---|---|---|---|---|---|
| 0.031(1) | 34.9 | 32.4 | 30.0 | 5.0 | 16.9 | - | 215.3 |
| 0.073(1) | 36.3 | 34.1 | 31.8 | 4.5 | 21.3 | - | 214.6 |
| 0.078(1) | 34.4 | 32.2 | 29.9 | 4.5 | 28.3 | - | 213.1 |
| 0.083(1) | 35.9 | 33.5 | 31.1 | 4.8 | 28.2 | - | 213.3 |
| 0.088(1) | 35.7 | 33.0 | 30.3 | 5.5 | 23.2 | - | 221.6 |
| 0.124(1) | 30.2 | 25.7 | 21.1 | 9.1 | 15.6 | - | 211.6 |
| 0.140(1) | 32.6 | 27.3 | 22.0 | 10.5 | 17.7 | - | 205.7 |
| 0.145(1) | 37.1 | 34.1 | 31.1 | 6.1 | 27.9 | - | 207.3 |
| 0.204(1) | 17.6 | - | - | - | - | - | 187.1 |
| 0.217(1) | 13.8 | - | - | - | - | - | 179.5 |
| 0.232(1) | 9.0 | - | - | - | - | - | 188.5 |
| 0.237(1) | 18.6 | - | - | - | - | - | 219.3 |
| 0.267(1) | - | - | - | - | - | 15.6 | 187.4 |
| 0.509(1) | - | - | - | - | - | 9.7 | - |
| 0.575(3) | - | - | - | - | - | p | - |
| 0.619(1) | - | - | - | - | - | 10.8 | - |
| 0.631(2) | - | - | - | - | - | 8.1 | - |
| 0.664(2) | - | - | - | - | - | 12.0 | - |